\documentclass{aastex62}
\usepackage{enumerate}

\shorttitle{CO(3--2) emission from the M31 disc}
\shortauthors{Li et al.}

\begin{document}

\title{The HASHTAG project I. A Survey of CO(3--2) Emission from the Star Forming Disc of M31}
\author{Zongnan Li}
\email{lzn@smail.nju.edu.cn}
\author{Zhiyuan Li}
\email{lizy@nju.edu.cn}
\affiliation{School of Astronomy and Space Science, Nanjing University, Nanjing 210023, China}
\affiliation{Key Laboratory of Modern Astronomy and Astrophysics, Nanjing University, Nanjing 210023, China}
\author{Matthew W. L. Smith}
\email{matthew.Smith@astro.cf.ac.uk}
\affiliation{School of Physics \& Astronomy, Cardiff University, The Parade, Cardiff CF24 3AA, UK}
\author{Christine D. Wilson}
\affiliation{Department of Physics and Astronomy, McMaster University, Hamilton, Ontario L8S 4M1, Canada}
\author{Yu Gao}
\affiliation{Purple Mountain Observatory and Key Laboratory for Radio Astronomy, Chinese Academy of Sciences, Nanjing 210023, China}
\author{Stephen A. Eales}
\affiliation{School of Physics \& Astronomy, Cardiff University, The Parade, Cardiff CF24 3AA, UK}
\author{Yiping Ao}
\affiliation{Purple Mountain Observatory and Key Laboratory for Radio Astronomy, Chinese Academy of Sciences, Nanjing 210023, China}
\author{Martin Bureau}
\affiliation{Sub-department of Astrophysics, University of Oxford, Denys Wilkinson Building, Keble Road, Oxford, OX1 3RH, UK}
\affiliation{Yonsei Frontier Lab and Department of Astronomy, Yonsei University, 50 Yonsei-ro, Seodaemun-gu, Seoul 03722, Republic of Korea}
\author{Aeree Chung}
\affiliation{Department of Astronomy, Yonsei University, 50 Yonsei-ro, Seodaemun-gu, Seoul 03722, Republic of Korea}
\author{Timothy A. Davis}
\affiliation{School of Physics \& Astronomy, Cardiff University, The Parade, Cardiff CF24 3AA, UK}
\author{Richard de Grijs}
\affiliation{Department of Physics and Astronomy, Macquarie University, Balaclava Road, Sydney NSW 2109, Australia}
\affiliation{Research Centre for Astronomy, Astrophysics and Astrophotonics, Macquarie University, Balaclava Road, Sydney NSW 2109, Australia}
\affiliation{International Space Science Institute--Beijing, 1 Nanertiao, Zhongguancun, Hai Dian District, Beijing 100190, China}
\author{David J. Eden}
\affiliation{Astrophysics Research Institute, Liverpool John Moores University, IC2, Liverpool Science Park, 146 Brownlow Hill, Liverpool L3 5RF, UK}
\author{Jinhua He}
\affiliation{Yunnan Observatories, Chinese Academy of Sciences, 396 Yangfangwang, Guandu District, Kunming, 650216, China}
\affiliation{Chinese Academy of Sciences South America Center for Astronomy, National Astronomical Observatories, Chinese Academy of Sciences, Beijing 100012, China}
\affiliation{Departamento de Astronom{\'i}a, Universidad de Chile, Casilla 36-D, Santiago, Chile}
\author{Tom M. Hughes}
\affiliation{Chinese Academy of Sciences South America Center for Astronomy, National Astronomical Observatories, Chinese Academy of Sciences, Beijing 100012, China}
\affiliation{Instituto de F\'{i}sica y Astronom\'{i}a, Universidad de
Valpara\'{i}so, Avda. Gran Breta\~{n}a 1111, Valpara\'{i}so, Chile}
\affiliation{CAS Key Laboratory for Research in Galaxies and Cosmology,
Department of Astronomy, University of Science and Technology of
China, Hefei 230026, China}
\affiliation{School of Astronomy and Space Science, University of Science and
Technology of China, Hefei 230026, China}
\author{Xuejian Jiang}
\affiliation{Purple Mountain Observatory and Key Laboratory for Radio Astronomy, Chinese Academy of Sciences, Nanjing 210023, China}
\author{Francisca Kemper}
\affiliation{European Southern Observatory, Karl-Schwarzschild-Strasse 2, Garching 85748, Germany}
\affiliation{Academia Sinica, Institute of Astronomy and Astrophysics, 11F Astronomy-Mathematics Building, NTU/AS campus, No. 1, Section 4, Roosevelt Rd., Taipei 10617, Taiwan}
\author{Isabella Lamperti}
\affiliation{Department of Physics and Astronomy, University College London, Gower Street, London, WC1E 6BT, UK}
\author{Bumhyun Lee}
\affiliation{Department of Astronomy, Yonsei University, 50 Yonsei-ro, Seodaemun-gu, Seoul 03722, Republic of Korea}
\author{Chien-Hsiu Lee}
\affiliation{National Optical Astronomy Observatory, 950 N. Cherry Avenue, Tucson, AZ 85719, USA}
\author{Micha\l$\;$J. Micha\l owski}
\affiliation{Astronomical Observatory Institute, Faculty of Physics, Adam Mickiewicz University, ul. S\l oneczna 36, 60-286 Pozna$\acute{n}$, Poland}
\author{Harriet Parsons}
\affiliation{East Asian Observatory, 660 N. A`oh$\bar{o}k\bar{u}$ Place, University Park, Hilo, HI 96720, USA}
\author{Sarah Ragan}
\affiliation{School of Physics \& Astronomy, Cardiff University, The Parade, Cardiff CF24 3AA, UK}
\author{Peter Scicluna}
\affiliation{Academia Sinica, Institute of Astronomy and Astrophysics, 11F Astronomy-Mathematics Building, NTU/AS campus, No. 1, Section 4, Roosevelt Rd., Taipei 10617, Taiwan}
\author{Yong Shi}
\affiliation{School of Astronomy and Space Science, Nanjing University, Nanjing 210023, China}
\affiliation{Key Laboratory of Modern Astronomy and Astrophysics, Nanjing University, Nanjing 210023, China}
\author{Xindi Tang}
\affiliation{Xinjiang Astronomical Observatory, Chinese Academy of Sciences, 830011 Urumqi, China}
\author{Neven Tomi$\rm \check{c}i\acute{c}$}
\affiliation{Max Planck Institute for Astronomy (MPIA), K$\ddot{o}$nigstuhl 17, 69117 Heidelberg, Germany}
\author{Sebastien Viaene}
\affiliation{Centre for Astrophysics Research, University of Hertfordshire, College Lane, Hatfield AL10 9AB, UK}
\affiliation{Sterrenkundig Observatorium, Universiteit Gent, Krijgslaan 281, 9000 Gent, Belgium}
\author{Thomas G. Williams}
\affiliation{School of Physics \& Astronomy, Cardiff University, The Parade, Cardiff CF24 3AA, UK}
\author{Ming Zhu}
\affiliation{National Astronomical Observatory of China, 20A Datun Road, Chaoyang District, Beijing 100012, China}

\begin{abstract}

We present a CO(3--2) survey of selected regions in the M31 disc as part of the JCMT large programme, HARP and SCUBA-2 High-Resolution Terahertz Andromeda Galaxy Survey (HASHTAG). The 12 CO(3--2) fields in this survey cover a total area of 60 square arcminutes, spanning a deprojected radial range of 2 -- 14 kpc across the M31 disc. Combining these observations with existing IRAM 30m CO(1--0) observations and JCMT CO(3--2) maps of the nuclear region of M31, as well as dust temperature and star formation rate surface density maps, we are able to explore the radial distribution of the CO(3--2)/CO(1--0) integrated intensity ratio ($R_{31}$) and its relationship with dust temperature and star formation. 
We find that the value of $R_{31}$ between 2 -- 9 kpc galactocentric radius is 0.14, significantly lower than what is seen in the nuclear ring at ~1 kpc ($R_{31}$ $\sim$ 0.8), only to rise again to 0.27 for the fields centred on the 10 kpc star forming ring. We also found that $R_{31}$ is positively correlated with dust temperature, with Spearman's rank correlation coefficient $\rho$ = 0.55. 
The correlation between star formation rate surface density and CO(3--2) intensity is much stronger than with CO(1--0), with $\rho$ = 0.54 compared to --0.05, suggesting that the CO(3--2) line traces warmer and denser star forming gas better. We also find that $R_{31}$ correlates well with star formation rate surface density, with $\rho$ = 0.69. 

\end{abstract}


\section{Introduction}

Star formation is one of the key processes in galaxy formation and evolution. 
As stars are born in cold gas, the rate of star formation is thought to scale with the total amount of cold gas (i.e. molecular plus atomic gas) 
known as the Kennicutt--Schmidt law \citep[KS law,][]{Schmidt 1959, Kennicutt 1989, Kennicutt et al. 2007}: $\Sigma_{\rm SFR} \propto \Sigma^\alpha_{\rm gas}$ with $\alpha$ $\approx$ 1.4, where $\Sigma_{\rm SFR}$ is the star formation rate (SFR) surface density and $\Sigma_{\rm gas}$ is the cold gas surface density. While atomic hydrogen can be directly detected through its 21 cm line, the H$_2$ molecule radiates weakly in rotational and vibrational states because it has no permanent electric dipole moment. Thus, the rotational transition of carbon monoxide (CO) is usually used as a proxy for H$_2$, as it is the second most abundant molecule, and its luminosity is approximately proportional to the total molecular gas mass \citep{Bolatto et al. 2013}.
\\

Recently, large surveys of nearby galaxies with advanced telescopes and instruments, including The HI Nearby Galaxy Survey \citep[THINGS;][]{Walter et al. 2008}, the Key Insights on Nearby Galaxies: A Far-Infrared Survey with Herschel \citep[KINGFISH;][]{Kennicutt et al. 2011}, the HERA CO-Line Extragalactic Survey \citep[HERACLES;][]{Leroy et al. 2009} and the JCMT Nearby Galaxies Legacy Survey \citep[NGLS;][]{Wilson 2012}, have enabled a detailed analysis of different gas components. These surveys revealed that the star-formation rate (SFR) surface density on and above a scale of a kiloparsec has a tighter correlation with the molecular gas surface density than with the total gas surface density, with a power-law index close to unity \citep[e.g.][]{Bigiel et al. 2008, Bigiel et al. 2011, Leroy 2008, Wilson 2012}. Despite the success on kpc scales, the behaviour of the KS law is still unclear below a scale of 500 pc, and a mystery below 100 pc \citep{Viaene 2018, Querejeta 2019}.
\\

Galaxies in the Local Group provide a unique laboratory to study the interstellar medium (ISM) and star formation, thanks to their proximity. In particular, at a distance of 780 kpc \citep{McConnachie 2005, de 2014}, where 1$\arcsec \sim$ 3.8 parsec, M31 provides a spatially resolved view of the ISM and star forming activities in a massive spiral. 
It is also a useful contrast to our own Galaxy because M31 has a more prominent classical bulge \citep[the classification of this bulge is still under debate, e.g.][]{Beaton 2007, Saglia 2018} and less prominent spiral arms than the Milky Way, with most of the current star formation occurring in the so-called 10 kpc ring \citep{Gordon et al. 2006}, which is likely a resonance phenomenon connected to the bulge and a weak bar \citep{Lewis et al. 2015}. 
\\

Previous surveys of CO(1--0) \citep{Nieten} and HI \citep{Braun et al. 2009} emission have revealed the global structure and kinematics of cold gas in M31. Observations of CO(2--1) lines toward several selected regions of M31 found the CO(2--1)/CO(1--0) intensity ratios to be 0.5 -- 0.7 in the arms \citep{Nieten} and close to unity in the central region \citep{Melchior 2011}, similar to the trend found in other disc galaxies. Despite the widespread use of CO(1--0) spectroscopy to trace the molecular gas, higher CO rotational transitions are needed to better constrain the molecular gas properties, in particular its excitation. The upper level energies of CO(1--0), CO(2--1), and CO(3--2) emission correspond to 5.5, 16.5 and 33.3 K, and the critical densities are $\sim 10^{3.5}$, $10^{4.3}$ and $10^{5}$ cm$^{-3}$, respectively \citep{Mao 2010}. In comparison, typical temperatures and densities of giant molecular clouds are $\sim$10 K \citep{Scoville 1987} and 10$^{2}$ -- 10$^3$ cm$^{-3}$ \citep{Solomon 1987}.  
Hence, compared with the two lower transitions of CO, CO(3--2) emission is a better tracer of warmer and denser gas that is more directly related to star formation \citep{Muraoka et al. 2007, Wilson 2009}. Furthermore, the CO(3--2)/CO(1--0) ratio can trace the variation of physical conditions of the bulk molecular gas, which, in fact, has been proved to be a good indicator of gas temperature \citep{Wilson et al. 1997} and density \citep[e.g.][]{Banerji 2009}.
\\

The dust temperature of M31 shows a general declining trend with galactocentric radius \citep[e.g.][]{Smith 2012}, and it is claimed that evolved stellar populations (a few Gyr old) are the primary heating source of dust, given that M31 is deficient in recent star formation activity \citep[e.g.][]{Montalto et al. 2009, Groves et al. 2012, Viaene et al. 2017}. Exploring the properties of the molecular gas in M31 could help determine whether M31 contains a large reservoir of very cold gas, and shed light on the relationship between gas and dust temperatures and the possible heating mechanisms.
\\ 

Taking advantage of the sensitive instruments on the James Clerk Maxwell Telescope (JCMT), we are carrying out the HARP and SCUBA-2 High-Resolution Terahertz Andromeda Galaxy Survey (HASHTAG), which consists of dust continuum observations at 450 and 850 $\mu$m, and spectroscopic observations of the CO(3--2) line. The full HASHTAG survey, which will include SCUBA-2 observations of the entire disc of M31, is described in the survey paper (M. Smith et al. in prep.). This paper presents initial results from the HASHTAG CO(3--2) survey, which is the first systematic census of CO(3--2) emission in M31. 
In Section 2, we describe the CO(3--2) observations and data reduction. In Section 3, we present products including CO(3--2) spectra, integrated intensity maps and the velocity field. We also investigate the CO(3--2)/CO(1--0) ratio in this section, based on which we address implications on the molecular gas properties. A summary is given in Section 4.
\\

\section{Observations and data reduction}

\subsection{HASHTAG CO(3--2) data}

The HASHTAG CO(3--2) observations were carried out with the 16-pixel array receiver HARP-B instrument \citep{Buckle 2009} and the Auto-Correlation Spectrometer Imaging System (ACSIS) configured to have a bandwidth of 1 GHz and a resolution of 0.488 MHz, corresponding to 0.43 km s$^{-1}$ at 345.796 GHz. The angular resolution of the JCMT at this frequency is 15$''$, with a pixel size of 7.5$''$ in the processed images. We have observed eleven 2$'\times 2'$ ($\sim$500 pc $\times$ 500 pc) jiggle fields (labelled a-k in Figure \ref{fig:1}) and one 4$'\times 4'$ (1 kpc $\times$ 1 kpc) raster field, together covering selected star forming regions across the M31 disc. All observations were taken between July and October, 2017. During this time two receptors were not operational, causing gaps in almost all jiggle fields, except for JIGGLEa and JIGGLEe, where the gaps were filled in owing to receiver rotation. The integration time was 3.5 hours for each of the 11 jiggle fields and 16.8 hours for the raster field in band 3 weather (0.08 $<\tau_{225}<$ 0.12, where $\tau_{225}$ represents the 225 GHz opacity) to achieve a sensitivity of 17 mK ($T\rm{_A^*}$, corrected antenna temperature) or better at a velocity resolution of 2.6 km s$^{-1}$. A log of the observations is given in Table \ref{tab:1}. \\

The footprints of the observed fields are shown in Figure \ref{fig:1}. The fields were selected based on the following considerations: 1) Five fields (JIGGLEa-e) overlap with regions observed by Herschel imaging in [CII] 158$\micron$, [OI] 63$\micron$ and [NII] 122$\micron$ and optical integral-field spectroscopy \citep{Kapala 2015}; 2) In two fields (JIGGLEf, g), it has been suggested that there is a component of very cold gas \citep{Allen 1993, Loinard 1996, Loinard 1998}; 3) Four additional fields (JIGGLEh-k) have been covered by multi-band optical imaging of the Panchromatic Hubble Andromeda Treasury \citep[PHAT,][]{Dalcanton 2012}, the CARMA CO(1--0) survey \citep[][Schruba et al. in prep.]{Caldu 2016} and IRAM CO(1--0)/CO(2--1) maps \citep{Nieten}, sampling a range of environments -- the bulge, the inner ring and the outer ring; 4) A 4$'\times4'$ region on the 10-kpc ring, which we observed in raster mode (half-array spacing). In total, these 12 fields cover an area of 60 square arcminutes, spanning a projected radial range of 2 -- 14 kpc across the M31 disc. 
\\

\begin{deluxetable*}{cccccccc}
\tablecaption{Observation Log}
\tablenum{1}
\tablewidth{2pt}
\tablehead{
\colhead{Field} &\colhead{RA (J2000)} &\colhead{Dec (J2000)}&
\colhead{PA} &\colhead{Coverage}&\colhead{Integration time}&
\colhead{RMS} & \colhead{$\tau_{225}$}\\
& & &
\colhead{($^\circ$)} &\colhead{($'$)}&\colhead{(hour)}&\colhead{(mK)} &
}
\decimalcolnumbers
\startdata
JIGGLEa & 00:46:31.0 & +42:11:51.5 &160.7&$2\times2$&3.5&16.7&0.09\\
JIGGLEb & 00:45:34.8 & +41:58:28.5  &145.7&$2\times2$&3.5&13.7&0.10\\
JIGGLEc &00:44:37.2 & +41:52:35.6&145.0 &$2\times2$&3.5&16.4&0.09\\
JIGGLEd &00:44:59.2 & +41:55:10.5&141.0 &$2\times2$&3.5&14.6&0.09\\
JIGGLEe & 00:44:26.5 & +41:37:12.7 & 153.0&$2\times2$&3.5&14.6&0.08\\
JIGGLEf & 00:43:03.3 & +41:24:16.2 & 130.0 &$2\times2$&3.5&15.9&0.10\\
JIGGLEg & 00:42:21.4 & +41:06:21.1 & 130.0 &$2\times2$&3.5&10.0&0.07\\
JIGGLEh &00:44:03.1 & +41:42:39.3 & 130.0 &$2\times2$&3.5&13.8&0.09\\
JIGGLEi & 00:44:13.2 & +41:35:17.1 & 130.0&$2\times2$&3.5&14.4&0.08\\
JIGGLEj & 00:45:26.9 & +41:44:54.6 & 37.7&$2\times2$&3.5&13.6&0.09\\
JIGGLEk & 00:43:52.2 & +41:33:48.9 & 37.7&$2\times2$&3.5&12.8&0.08\\
Raster & 00:44:40.9 & +41:27:25.2 & 37.7&$4\times4$&16.8&14.5&0.08\\
\enddata
\tablecomments{(1) Field name. (2)--(4) Centres and position angles of these fields. (5) Coverage of the fields. (6) Total integration time of each field. (7) Mean RMS noise in $T\rm{_A^*}$, measured at a velocity resolution of 2.6 km s$^{-1}$. (8) The 225 GHz opacity, representing the weather conditions. 
\label{tab:1}}
\end{deluxetable*}

The data reduction follows the standard procedure for JCMT heterodyne observations, using the ORAC-DR pipeline \citep{Jenness 2015} based on the $Starlink$ software\footnote{http://starlink.eao.hawaii.edu} \citep{Currie 2014}. The pipeline automatically flagged spikes and bad detectors, subtracted linear baselines and assessed data quality. However, after both automatically and manually flagging spectra with bad baselines, we noticed that one receptor remained questionable, which caused humps on the baselines in different fields. As the lines in our fields are relatively narrow ($\sim$ 30 km s$^{-1}$) compared to the whole bandwidth ($\sim$ 800 km s$^{-1}$), and the humps are relatively far away from the signal, we truncated the spectra down to 200 km s$^{-1}$ wide to avoid the baseline noise caused by the humps. \\

We followed the methods described by \cite{Wilson 2012} to obtain the moment maps. The main steps are summarized below. First, we shifted the velocity reference frame from HELIOCENTRIC (with respect to the sun) to kinematical Local Standard of Rest (LSRK) and divided each spectrum by the main-beam efficiency ($\eta\rm_{MB}$) 0.64 to convert from antenna temperature to main-beam temperature. Then, we smoothed the CO(3--2) data cube to an angular resolution of 23$\arcsec$, 
corresponding to 88 pc in projection, and regridded the velocity resolution to 2.6 km s$^{-1}$ 
to match the CO(1--0) survey data \citep[][Section 2.2]{Nieten}. Next, since the signal-to-noise ratio (S/N) of different fields varies due to different system temperatures, integration times and unstable receptors, we divided the original data cubes by the noise maps produced using the line-free channels of each field to obtain the S/N cube. 
\\

We then applied the classical ClumpFind algorithm \citep{Williams 1994} implemented as part of the $Starlink$/CUPID task $findclumps$ to the CO(3--2) data. By contouring the data and searching for the local maximum in the three-dimensional S/N cubes, this algorithm identifies clumps with emission above three times the RMS noise and containing at least 50 pixels, with a pixel size of 7.5$\arcsec$ (corresponding to $\sim$30 pc). Pixels that do not belong to any of the clumps are subsequently masked. 
Two (five) times the RMS noise was applied to the JIGGLEg and h (Raster) fields, since their signals are relatively weaker (stronger). The typical size (defined as the full width at half maximum; FWHM) of the clumps is $\sim$90 pc, and the typical line width is 9 km s$^{-1}$. A list of the resultant clumps is given in the Appendix. 
\\

Taking advantage of the $findclumps$ procedure, blended components along the line-of-sight were disentangled and identified as different clumps. Application of this procedure has also improved the significance of the signal by eliminating the patchy local maxima that were much smaller than beam size which are most likely statistical fluctuations. Although there are a variety of clump finding algorithms that are also included in the CUPID package \citep[GAUSSCLUMP, REINHOLD and FELLWALKER][see \cite{Watson 2010} for a comprehensive assessment]{}, 
the choice of specific algorithm does not significantly affect our conclusion, since our main goal is to find potentially coherent regions with sufficient S/N for analysis and comparison, but not to carry out a rigorous identification of individual molecular clouds. 
Moreover, we note that the detailed parameter selection of this algorithm has little influence on our results, as lowering the S/N cutoff or the pixel threshold can only add a few more clumps in total.
Moment maps were created from the original data cube using the mask produced by $findclumps$. 
\\

\subsection{Ancillary data}

We used the CO(1--0) map from the IRAM 30m survey presented by \cite{Nieten}, which fully sampled an area of 2$^{\circ}\times 0.5^{\circ}$ with an angular resolution of 23$''$. The mean RMS noise of the CO(1--0) integrated intensity map is 0.35 K km s$^{-1}$. We truncated the CO(1--0) image to match the fields of our observations using the $imregrid$ task in CASA \citep{McMullin 2007}, except for JIGGLEa which is at the far side of the disc and was not covered by the CO(1--0) survey. We then applied the same procedure as described in Section 2.1 to identify clumps and obtain moment maps of CO(1--0). 
Here we did not include the CO(2--1) data from the same survey, since those observations do not completely match our CO(3--2) fields, and the data are currently not available to us. 
\\

\begin{figure}
\centering
\includegraphics[width=6in]{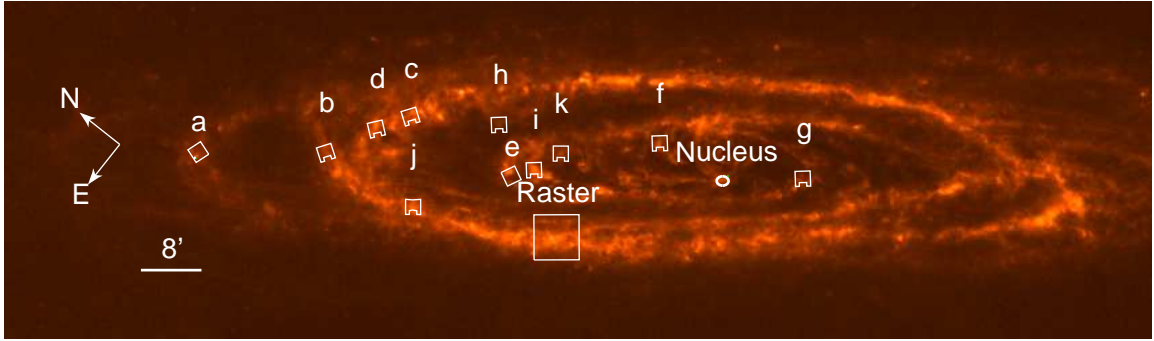}\\
\caption{The footprint of the 11 HASHTAG jiggle fields (small white squares) and one raster field (large white box) overlaid on the Herschel/SPIRE 250 $\mu$m image of M31. The small ellipse labelled `Nucleus' indicates a region on the nuclear ring studied by \cite{Li et al. 2019}. 
\label{fig:1}}
\end{figure}

We also made use of a CO(3--2) map of the central $\sim$ 1 kpc radius of M31 obtained in JCMT HARP raster mode \citep[see the ellipse in Figure \ref{fig:1},][]{Li et al. 2019}, taken with a resolution of 15$''$ and a typical RMS of 3.5 mK in a 13 km s$^{-1}$ channel. Due to the general deficiency of molecular gas in this region, only a small portion of the so-called nuclear ring covered by our field-of-view has CO(1--0) detections, from which we can derive the CO(3--2)/CO(1--0) ratios \citep{Li et al. 2019}. Due to the shallower effective integration, the CO(3--2) RMS here is higher ($\sim$25 mK). These data complement the HASHTAG observations in terms of radial coverage. \\

The dust temperature map used here is from \cite{Smith 2012} (Figure \ref{fig:1}), which is based on spectral energy distribution fitting of $Herschel$ Photodetector Array Camera and Spectrometer (PACS) and Spectral and Photometric Imaging Receiver (SPIRE) 100, 160, 250, 350, and 500 $\mu$m observations along with $Spitzer$ Multiband Imaging Photometer (MIPS) 70 $\mu$m data. A Bayesian method \citep[PPMAP,][]{Marsh 2015} has been applied to increase the angular resolution of this map to 8$\arcsec$ \citep{Whitworth 2019}. 
\\

The SFR surface density map with an angular resolution of 6$\arcsec$ is from \cite{Ford 2013}\footnote{In order to compare the inclination corrected SFR density map with our uncorrected CO data, we divided this map by cos $i$, with the inclination angle of M31, $i=77^\circ$ \citep{McConnachie 2005}, which was applied by \cite{Ford 2013}.}, which is calibrated based on the $GALEX$ far-ultraviolet (FUV) map of M31 \citep{Thilker et al. 2005} and a $Spitzer$ MIPS 24 $\mu$m \citep{Gordon et al. 2006} map, using the method prescribed by \cite{Leroy 2008}. \cite{Ford 2013} also used $GALEX$ near-ultraviolet (NUV) and $Spitzer$ Infrared Array Camera (IRAC) 3.6 $\mu$m maps to correct for foreground stars and emission from old stellar populations, respectively. Recently, \cite{Lewis 2017}, using the PHAT data, suggested that the FUV plus 24 $\mu$m method underestimates the SFR by a factor of 2.3 -- 2.5. Furthermore, \cite{Tomicic 2019} used the extinction-corrected H$\alpha$ line of five kpc-sized fields in M31 (coinciding with the positions of the JIGGLE a-e fields) to calibrate the SFR, and claimed it is 5 times higher than derived by \cite{Ford 2013}. For consistency of cross-field comparison, we still adopt the SFR map from \cite{Ford 2013}, bearing in mind that in most of our fields the actual SFR may potentially shift up by $\sim$0.5 dex.
\\

\section{Results}

In this section, we focus on the CO(3--2) line measurements, in a close comparison with the CO(1--0) emission observed by the IRAM survey. We also address the immediate implications for the physical conditions of the molecular gas and star formation in the disc of M31.

\subsection{Line characteristics and morphology of the molecular gas}

\begin{figure}
\centering
\includegraphics[width=3.5in]{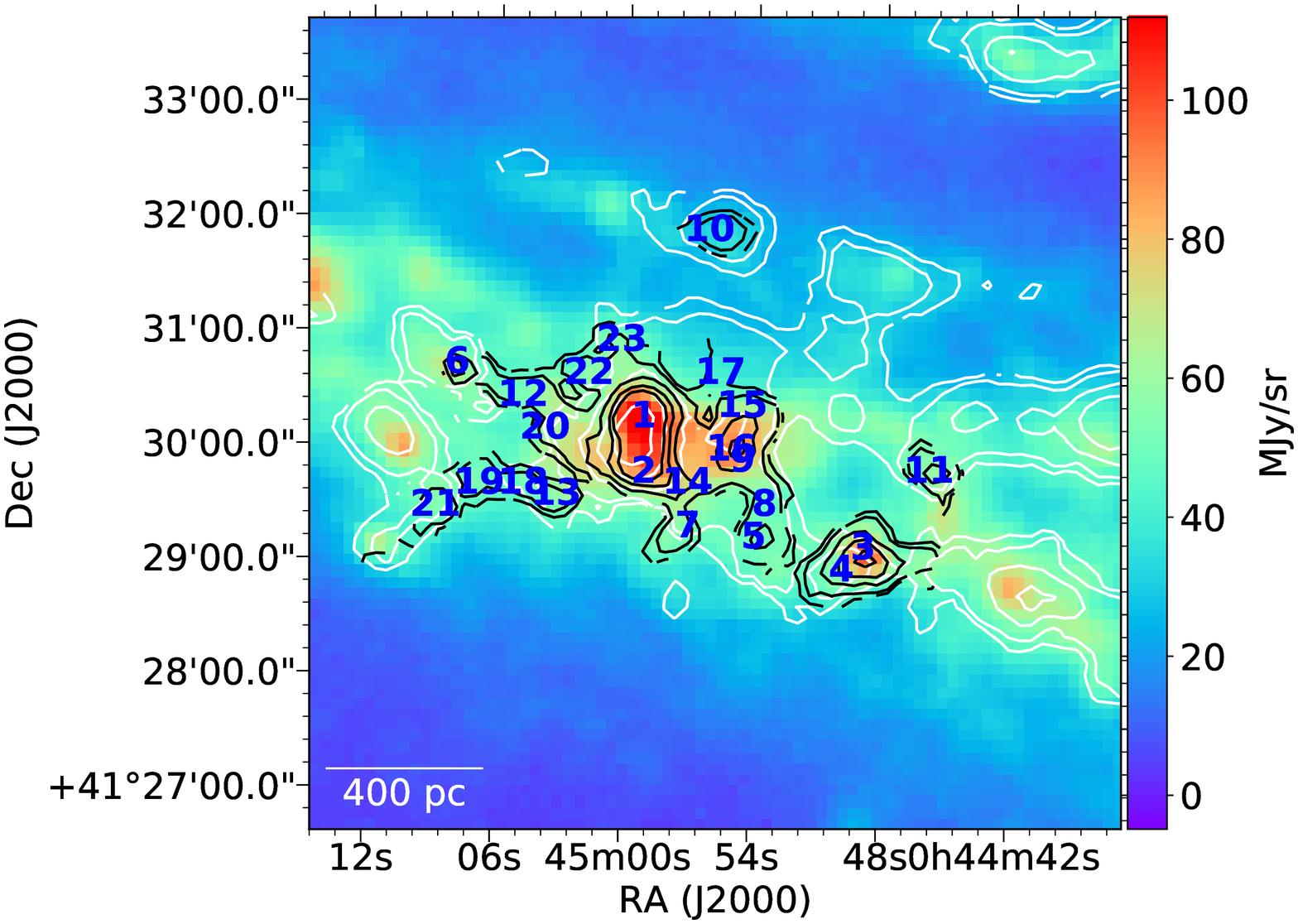}
\includegraphics[width=3.5in]{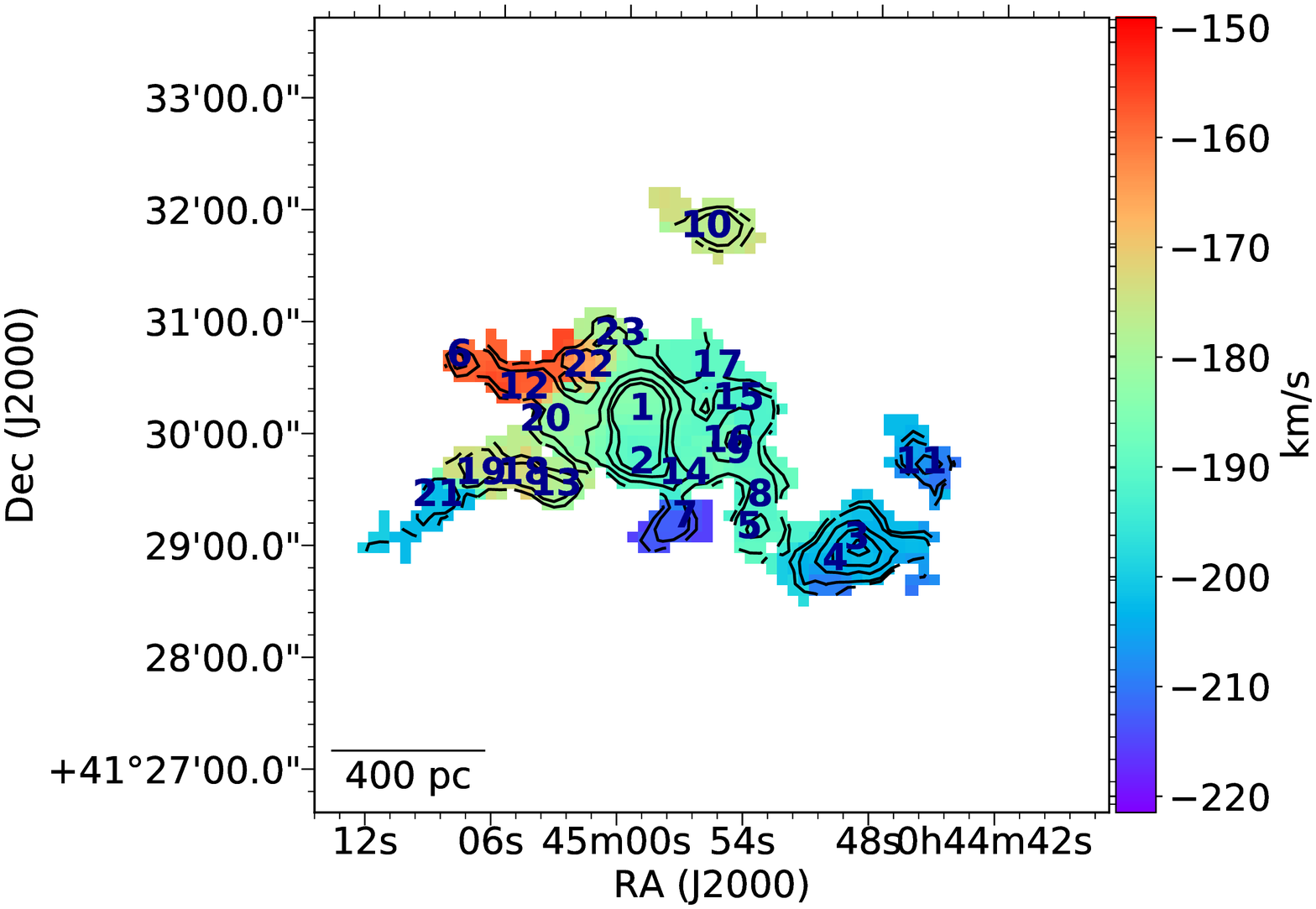}
\caption{$Left$: Herschel/SPIRE 250 $\mu$m maps of the raster field overlaid with CO(1--0) (white) and CO(3--2) integrated intensity contours (black), with contour levels as in Figure 2. $Right$: The CO(3--2) velocity field map of the raster field, overlaid with CO(3--2) contours. The clumps identified by ClumpFind are labelled with numbers. \label{fig:3}}
\end{figure}

The integrated intensity contours of CO(3--2) (black) and CO(1--0) (white) are contrasted with the $Herschel$/SPIRE 250 $\mu$m map in Figures \ref{fig:2} and \ref{fig:3}. Identified clumps are labelled in these maps. In some cases, such as clumps 1 and 2 in the JIGGLEb field, 
two components superposed along the line-of-sight can be separated by their velocity channels. Since the clumps are labelled at their barycenters (i.e., values given in Table \ref{tab:3}) rather than their peaks, some clumps with peculiar shapes may appear to deviate from the strongest peaks (e.g. in the JIGGLEi field). 
The CO(3--2) emission is in general coincident with the CO(1--0) and dust emission, with CO(3--2) being more compact. In some fields, such as JIGGLEf, g and h, the CO(3--2) emission is relatively weak, and does not trace the dust emission well. This is understandable, because the JIGGLEf and g fields are suggested to contain very cold gas \citep{Allen 1993, Loinard 1996, Loinard 1998}, and JIGGLEh samples the interarm region, where molecular gas may be scarce and star formation is inactive. It is also possible that the CO clump is aligned with the dust continuum that is not associated with it. The high inclination of M31 can project clouds in such a way that they appear close to each other.
\\

The CO(3--2) line-of-sight velocity map of the Raster field is shown in the right-hand panel of Figure \ref{fig:3}. It is in agreement with the CO(1--0) velocity field and consistent with the rotation pattern of the M31 disc. The two blue-shifted patches at the lower left (labelled `7' and `21') seem inconsistent with the bulk rotation pattern, which could be due to projection of clouds at different distances or outflows with velocities on the order of tens of km s$^{-1}$ on a scale of $\sim$ 100 pc. It will be interesting to explore the nature of these anomalous regions in future work.
\\

\begin{figure}
\centering
\includegraphics[width=6in]{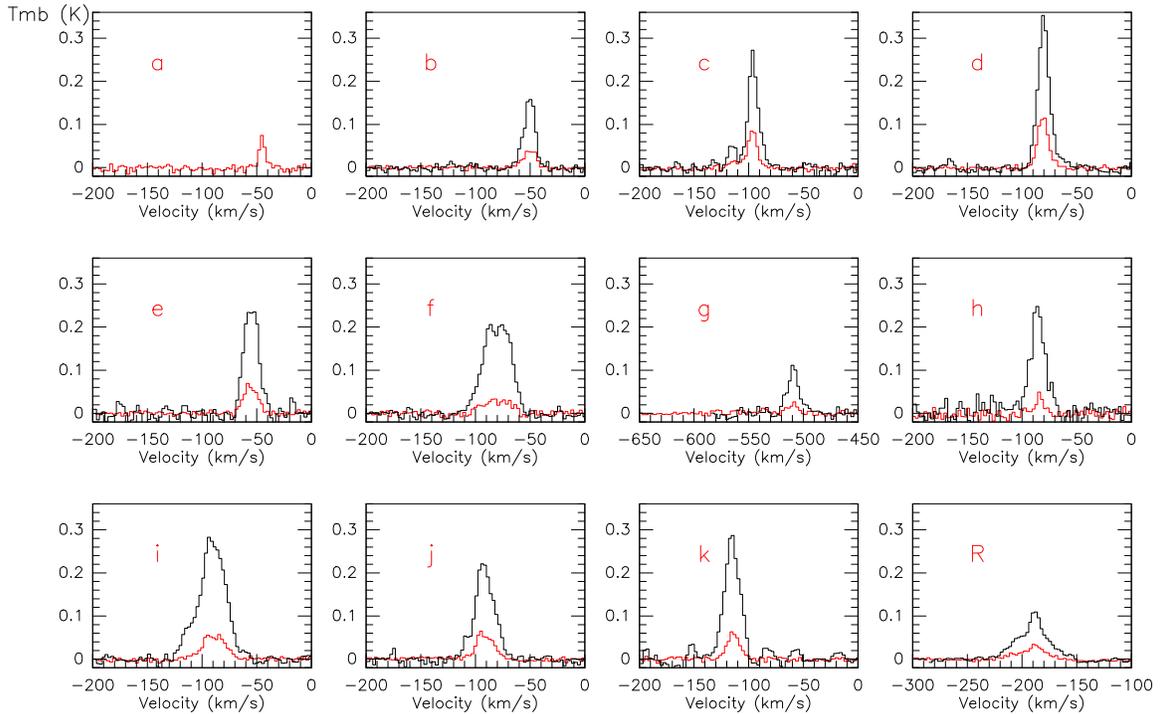}
\caption{The CO(3--2) (red) and CO(1--0) (black) line profiles of the 12 fields, averaged using CO(3--2) spectra with peak S/N $>$ 5 and CO(1--0) spectra from the same region. Field `a' was not covered in CO(1--0) by the survey of \cite{Nieten}.\label{fig:4}}
\end{figure}

For each field, we stacked the CO(3--2) spectra with peak S/N greater than 5 and CO(1--0) spectra from the same region as CO(3--2), and then divided them by the number of spectra. The averaged spectra of CO(3--2) (red) and CO(1--0) (black) of each field are shown in Figure \ref{fig:4}. Except for the JIGGLEa field without CO(1--0) observations, the two lines of the other 11 fields agree well with each other, both in central velocity and velocity dispersion (see also the Appendix). The presence of broad lines and multiple components in some fields indicates blending of clumps either along the line-of-sight or across the field-of-view. Despite this general agreement, there are a few fields with additional CO(1--0) components (JIGGLEc, e, k) or much broader CO(1--0) lines (JIGGLEh, i and the Raster field). 
This could be the case if the corresponding CO(3--2) emission is too weak or absent along the line-of-sight. It is clear that in all fields the CO(3--2) lines are much weaker than the CO(1--0) lines. 
\\

\subsection{CO(3--2)/CO(1--0) line ratios and gas properties}

The CO(3--2)/CO(1--0) line ratio, $R_{31}$, has been suggested as a good indicator of temperature in the molecular gas, with higher ratios corresponding to higher temperatures \citep{Wilson et al. 1997}. It has also been suggested as an indicator of gas density \citep[e.g.][]{Banerji 2009}. Typical $R_{31}$ values are found to be 0.4--0.5 in the disc of the Milky Way \citep{Sanders et al. 1993, Oka et al. 2007}, and similar values have been found in normal spiral galaxies, such as in NGC 4254 and NGC 4321 \citep{Wilson 2009}, M51 \citep{Vlahakis et al. 2013}, and NGC 628 \citep{Muraoka 2016}. 
Comparison with these values could provide us with a deeper insight into the physical conditions of the molecular gas in M31.
\\

To distinguish possible multiple components along the line-of-sight, we produced moment maps of individual clumps found by the ClumpFind algorithm as described in Section 2.1, from which we derived the integrated intensity, central velocity and velocity dispersion of each spatial pixel in a given clump. We then identified the CO(3--2) and CO(1--0) lines of each pixel arising from the same clump if the difference of their central velocities is less than two channel widths (5.2 km s$^{-1}$) and both lines have velocity dispersions greater than one channel width (2.6 km s$^{-1}$). This selection excluded $\sim$60\% of pixels with CO(1--0) detection but no CO(3--2) detection and $\sim$7\% of pixels with CO(3--2) detection but no CO(1--0) detection, which is expected since CO(3--2) tends to probe the denser regions of molecular clouds. We provide the pixel-by-pixel velocity information of the two lines in the Appendix. The following correlation analyses are based on the pixels thus selected. We note that these pixels are not fully independent. However, this should have little effect on our results as far as a global correlation is concerned. 
\\

\begin{figure}
\centering
\includegraphics[width=5in]{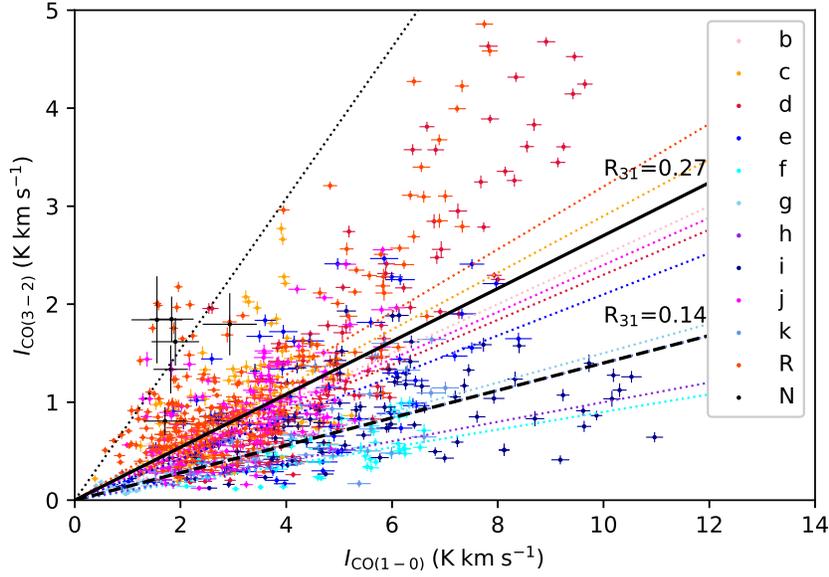}\\
\caption{The pixel-by-pixel correlation of the integrated intensity of the 12 HASHTAG fields, plus the additional Nucleus field (denoted by `N') studied by \cite{Li et al. 2019}. Each pixel is obtained from the moment maps based on individual clumps found by ClumpFind and selected using the criteria described in Section 2.1. Only the measurement uncertainties are shown in the plot. 
The red-tinted points represent the fields in the 10 kpc ring, with the solid line indicating their mean CO(3--2)/CO(1--0) line ratio, 0.27, while the blue-tinted points are from the other fields in the disc, with the dashed line representing their mean line ratio of 0.14. The colour-coded dotted lines indicate the mean line ratios for each field, as listed in Table \ref{tab:2}.}\label{fig:5}
\end{figure}

Figure \ref{fig:5} shows the pixel-by-pixel correlation of integrated intensities of the two lines. The measurement uncertainties in this plot are given by $\Delta T \times \Delta V \times \sqrt{N\rm_{chan}}$ \citep{Wilson 1989}, where $\Delta T$ is the baseline RMS of each pixel, $\Delta V$ the channel width, and $N\rm_{chan}$ the number of channels used to produce the moment 0 maps. The uncertainties range from $\sim$1\% -- 30\% for CO(3--2) and $\sim$1\% -- 20\% for CO(1--0) in all fields, with an average value of $\sim$10\% and 5\%, respectively. The calibration uncertainties for CO(1--0) and CO(3--2) observations are not included here, which are 8\% \citep{Saintonge 2017} and 10\%\footnote{https://www.eaobservatory.org/jcmt/instrumentation/heterodyne/calibration/}, respectively. They are taken into account in the line ratios calculated below. 
\\

\begin{figure}
\centering
\includegraphics[width=2.3in]{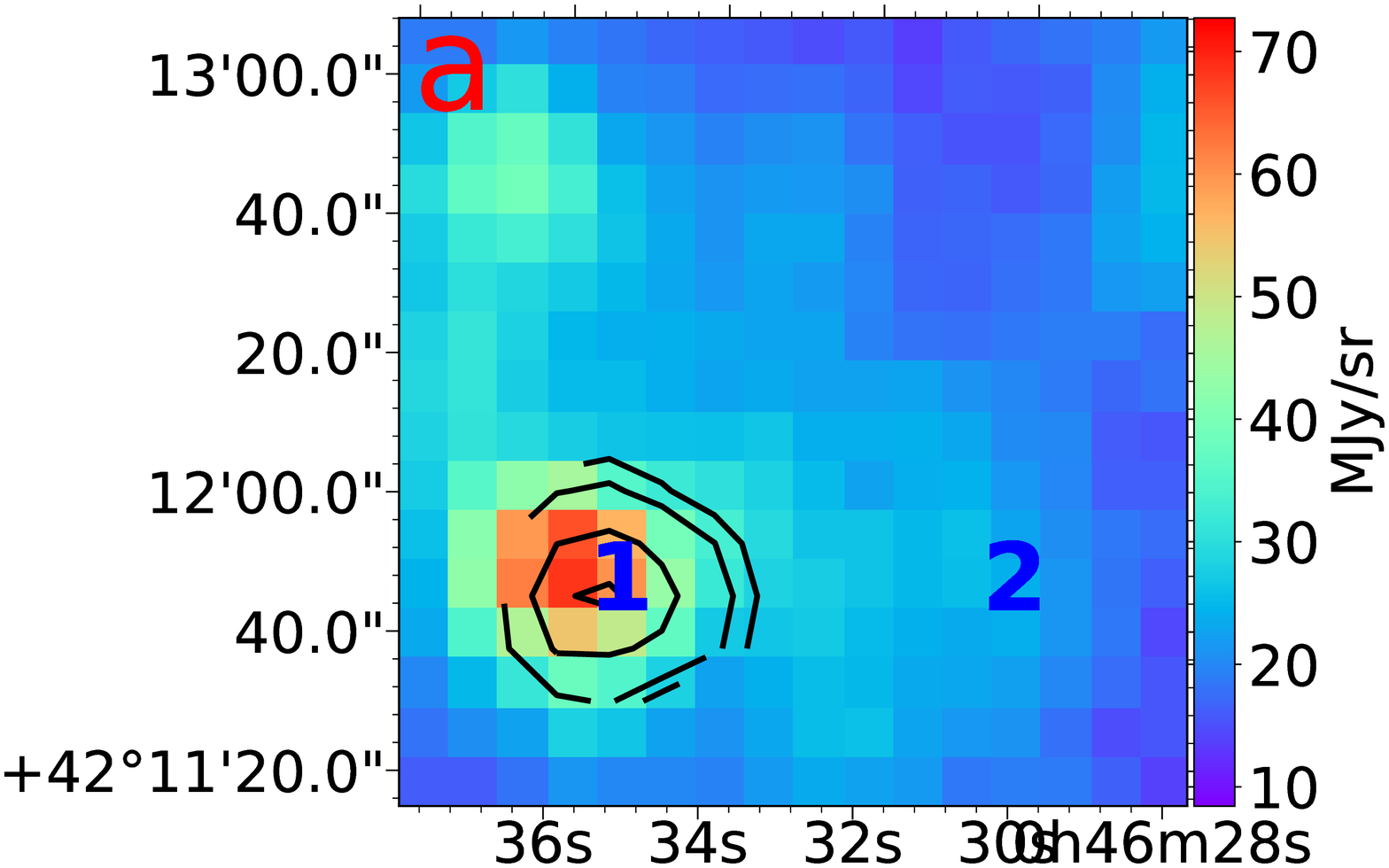}
\includegraphics[width=2.3in]{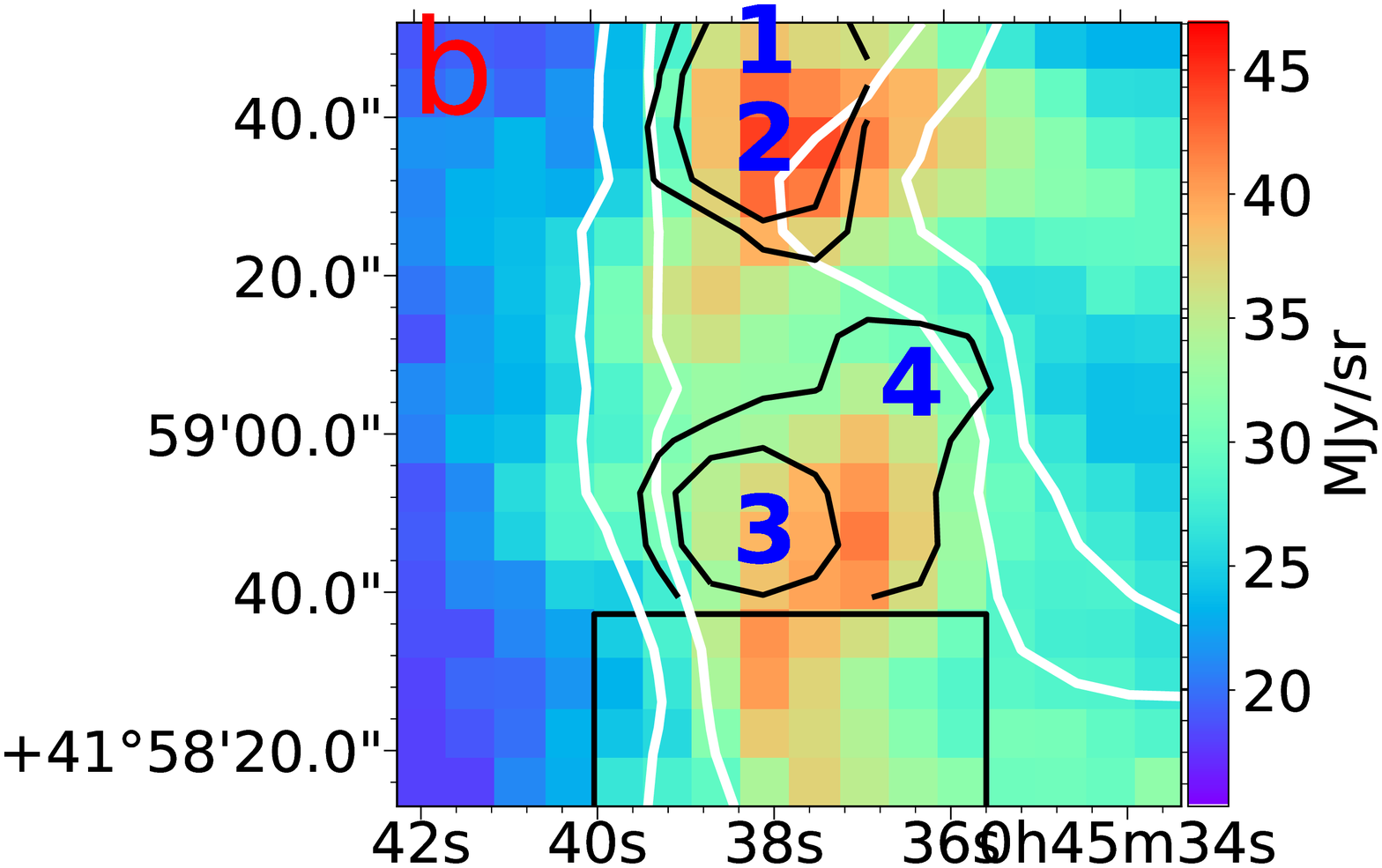}
\includegraphics[width=2.3in]{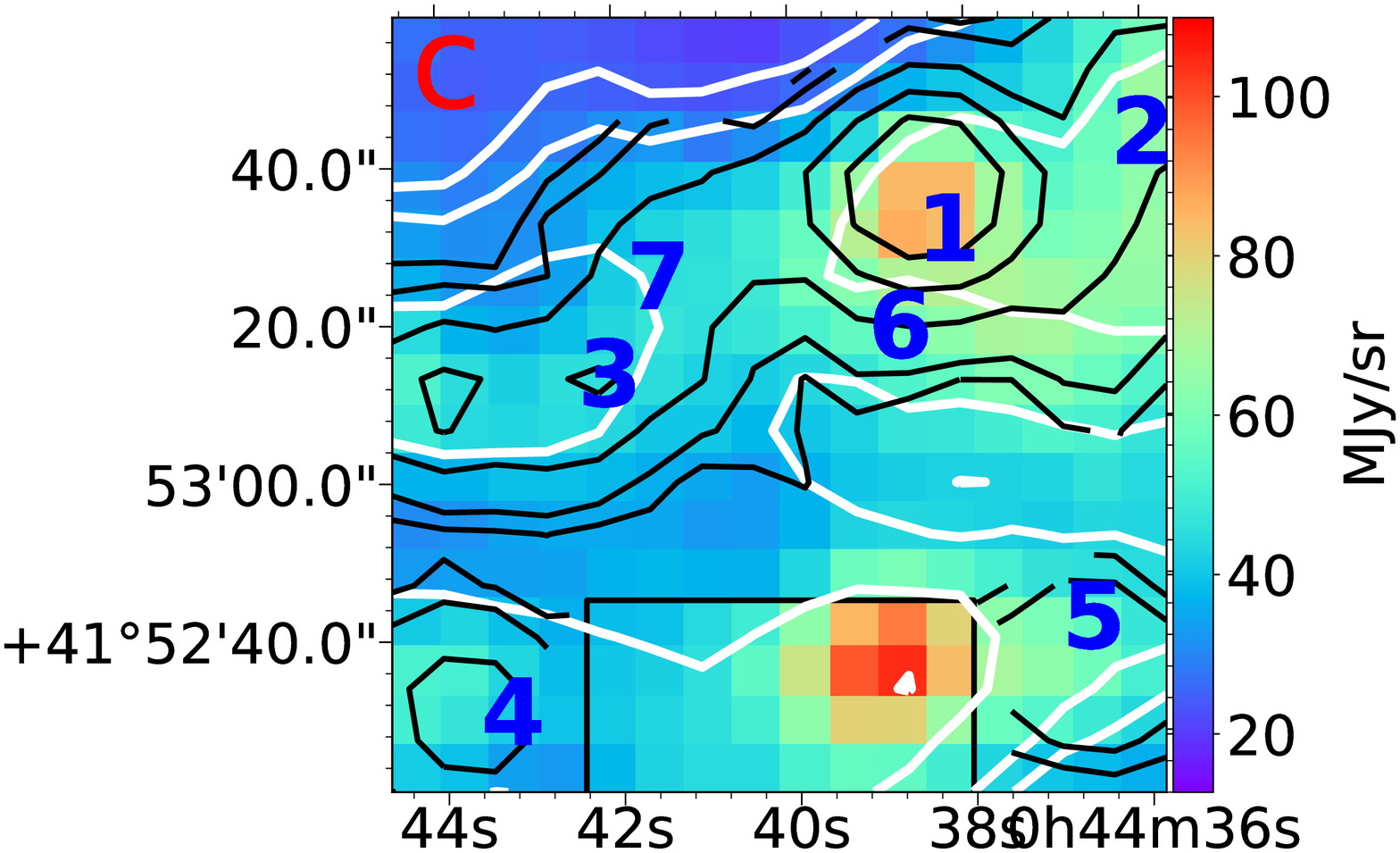}
\includegraphics[width=2.3in]{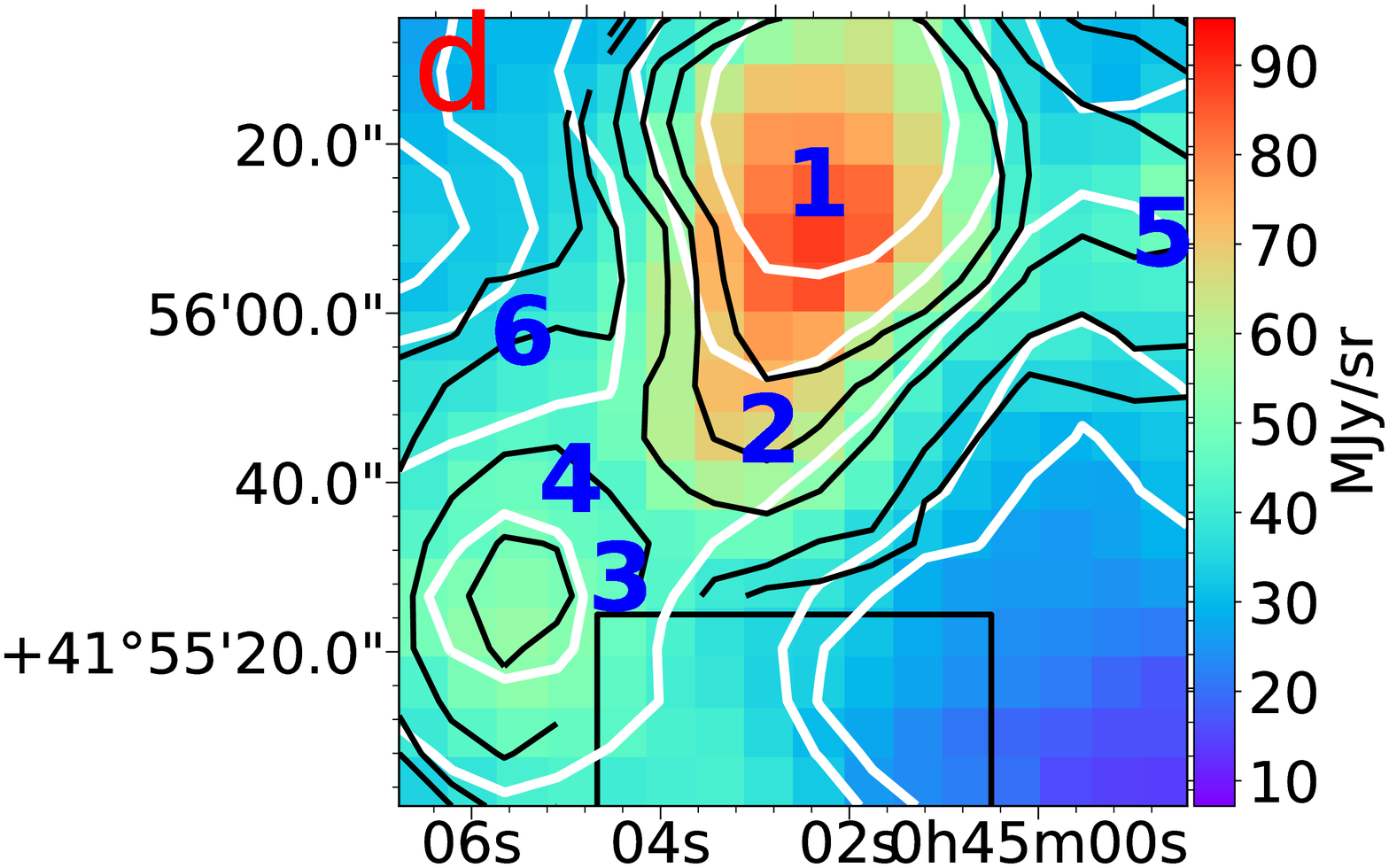}
\includegraphics[width=2.3in]{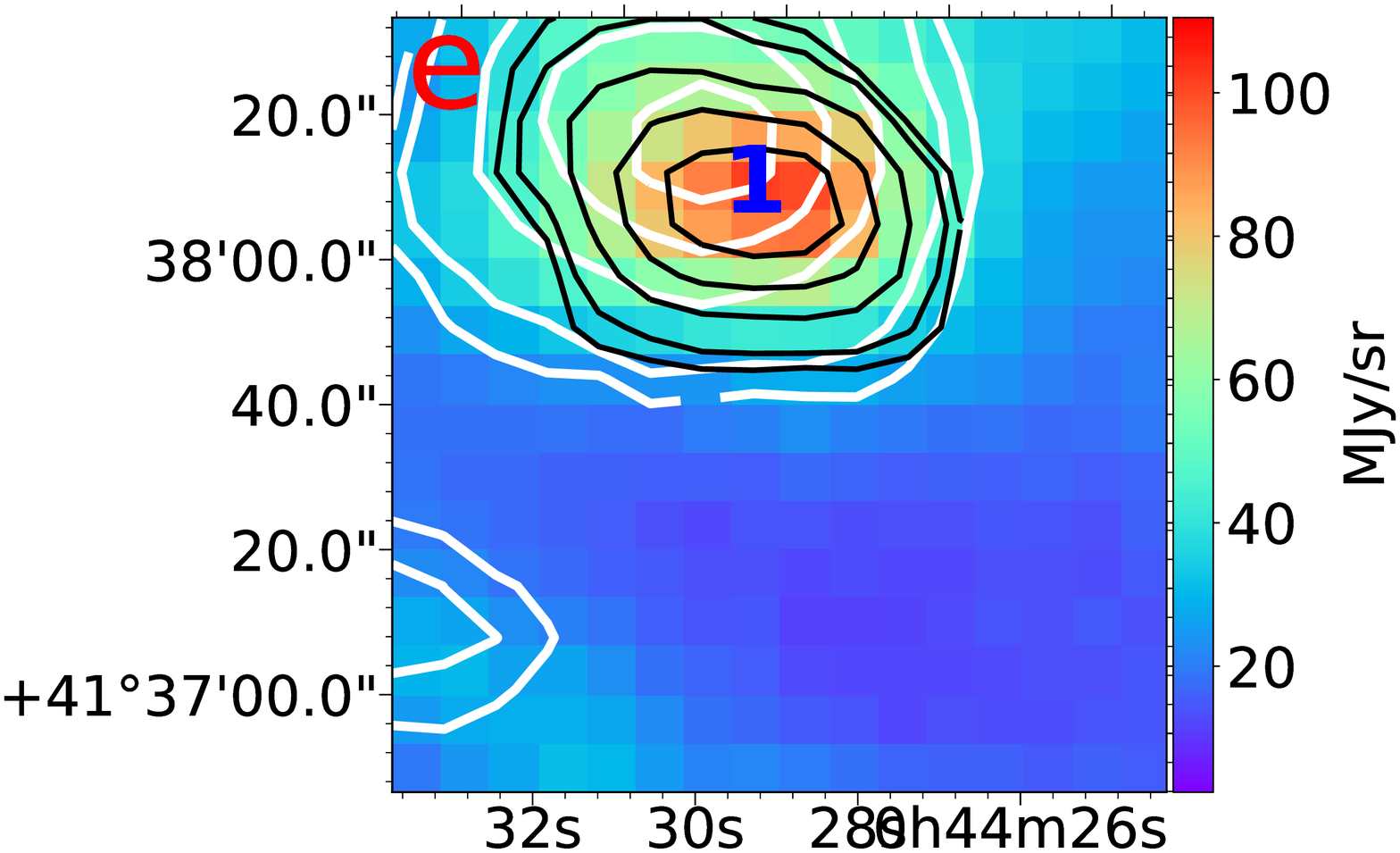}
\includegraphics[width=2.3in]{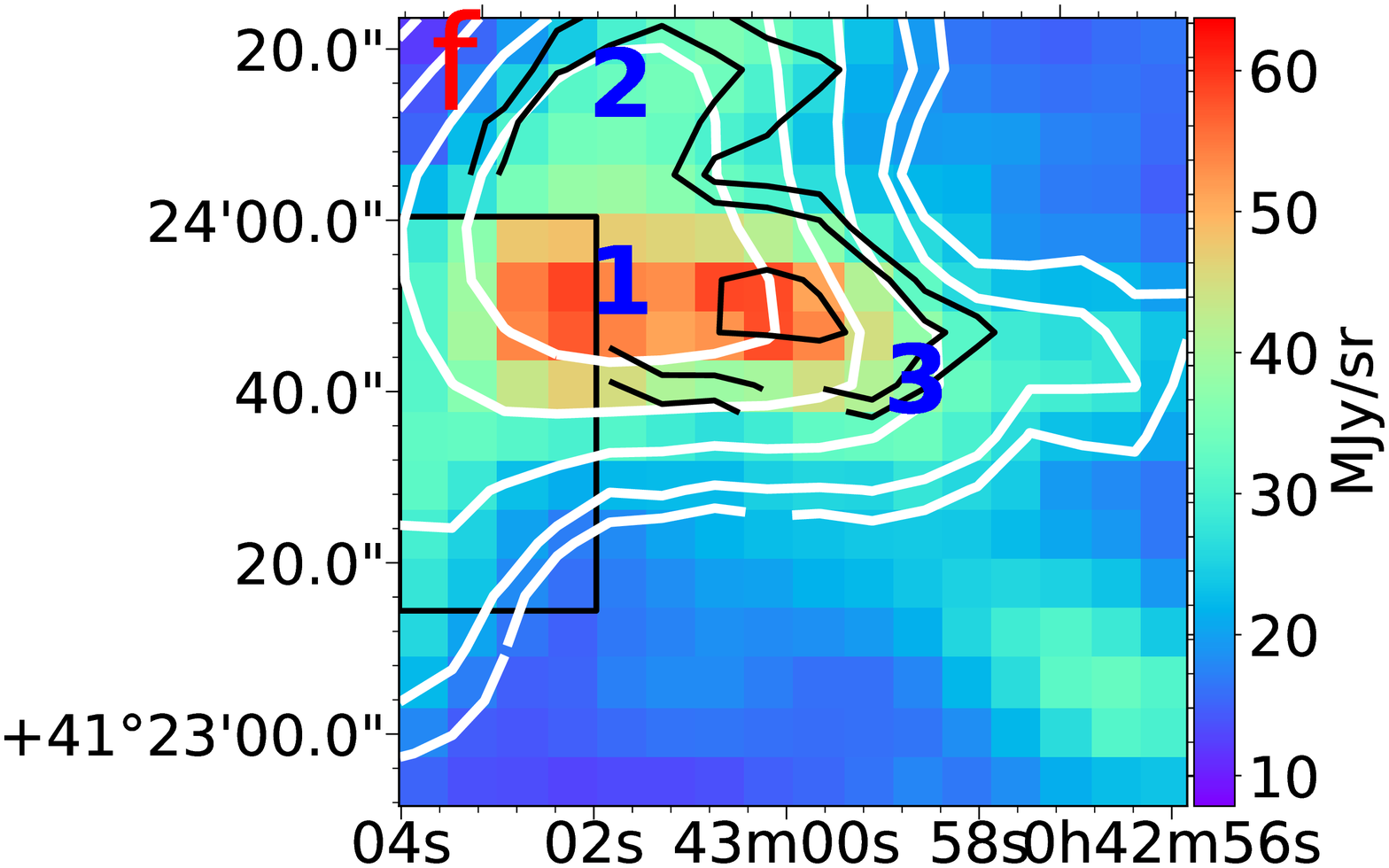}
\includegraphics[width=2.3in]{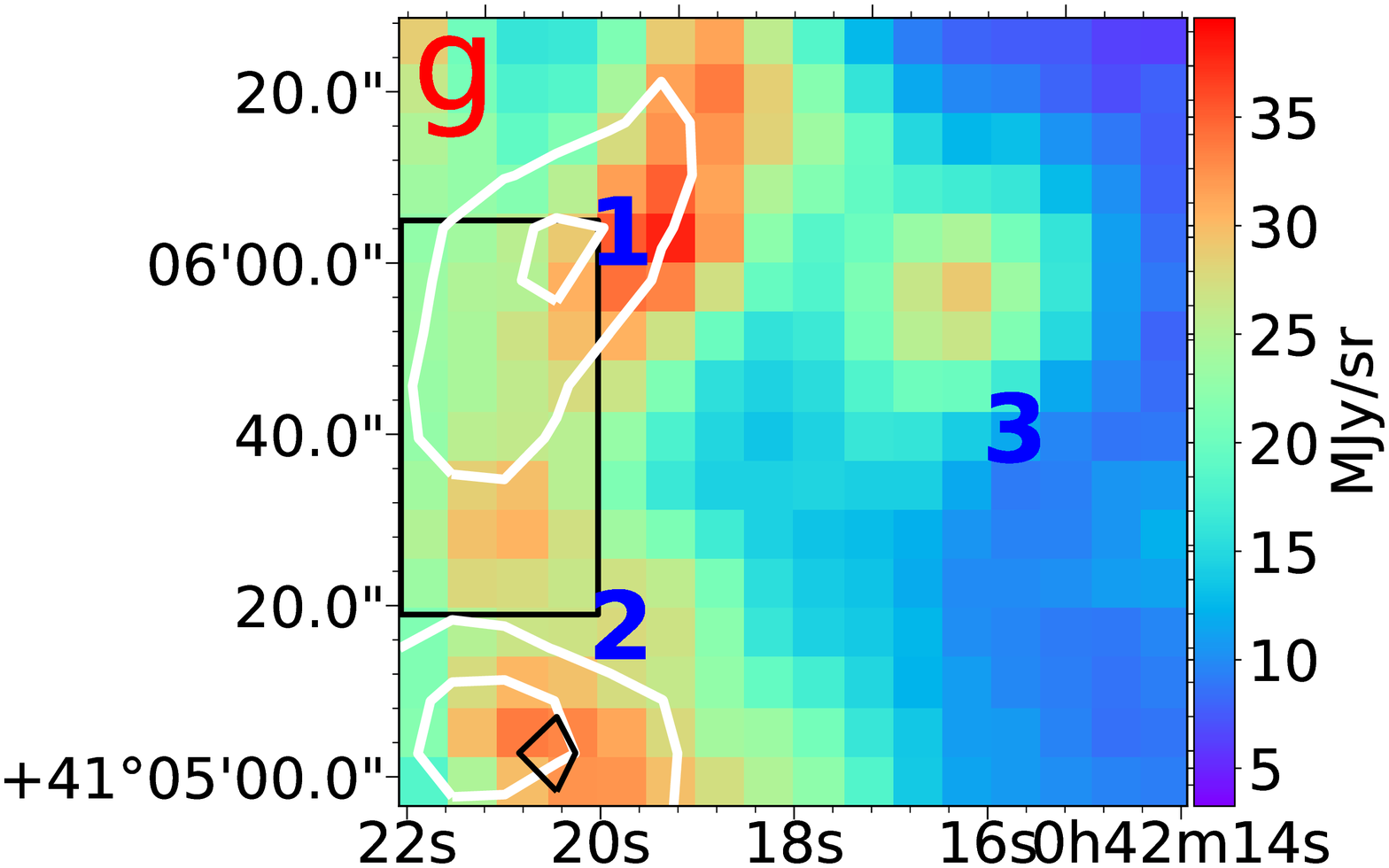}
\includegraphics[width=2.3in]{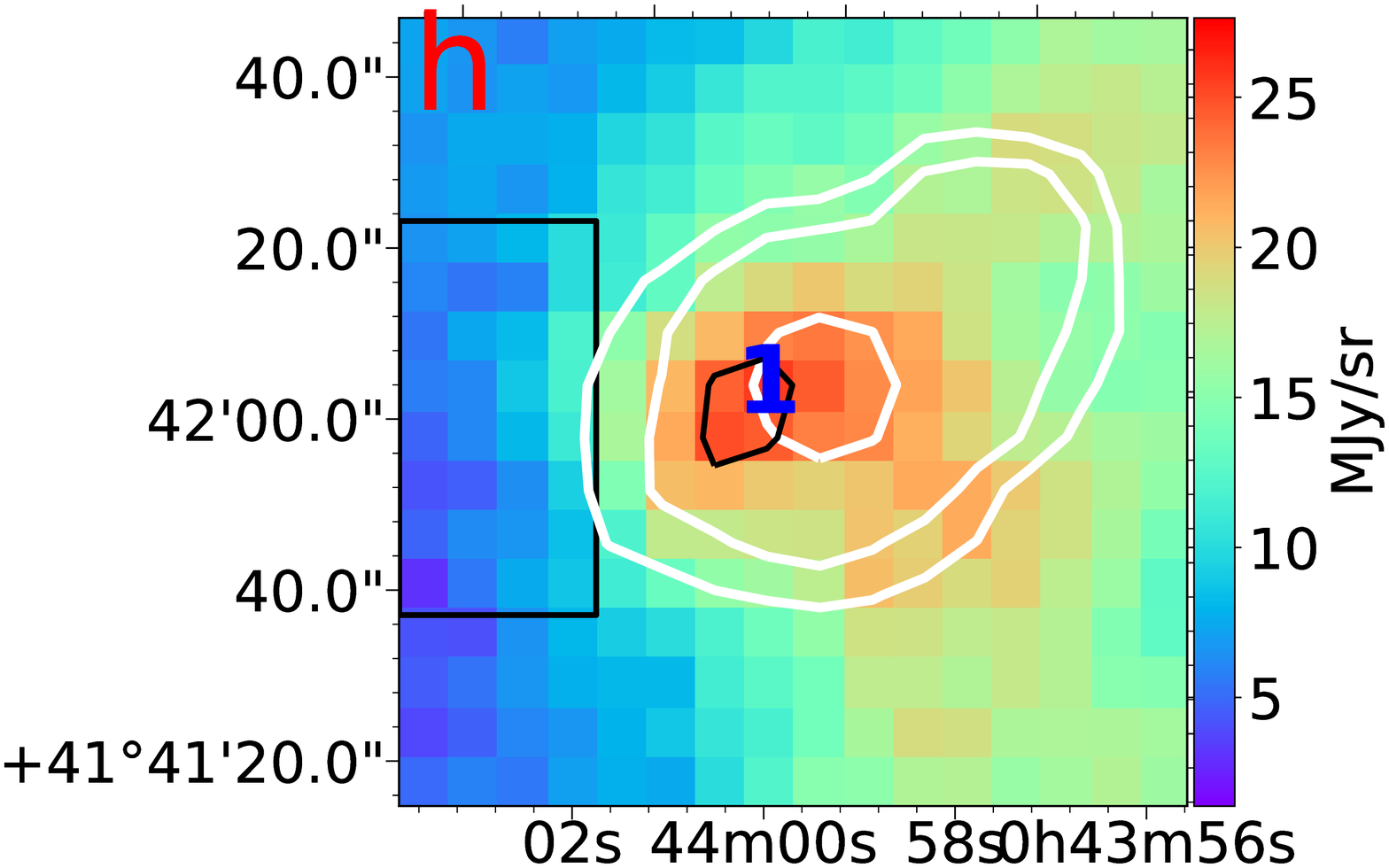}
\includegraphics[width=2.3in]{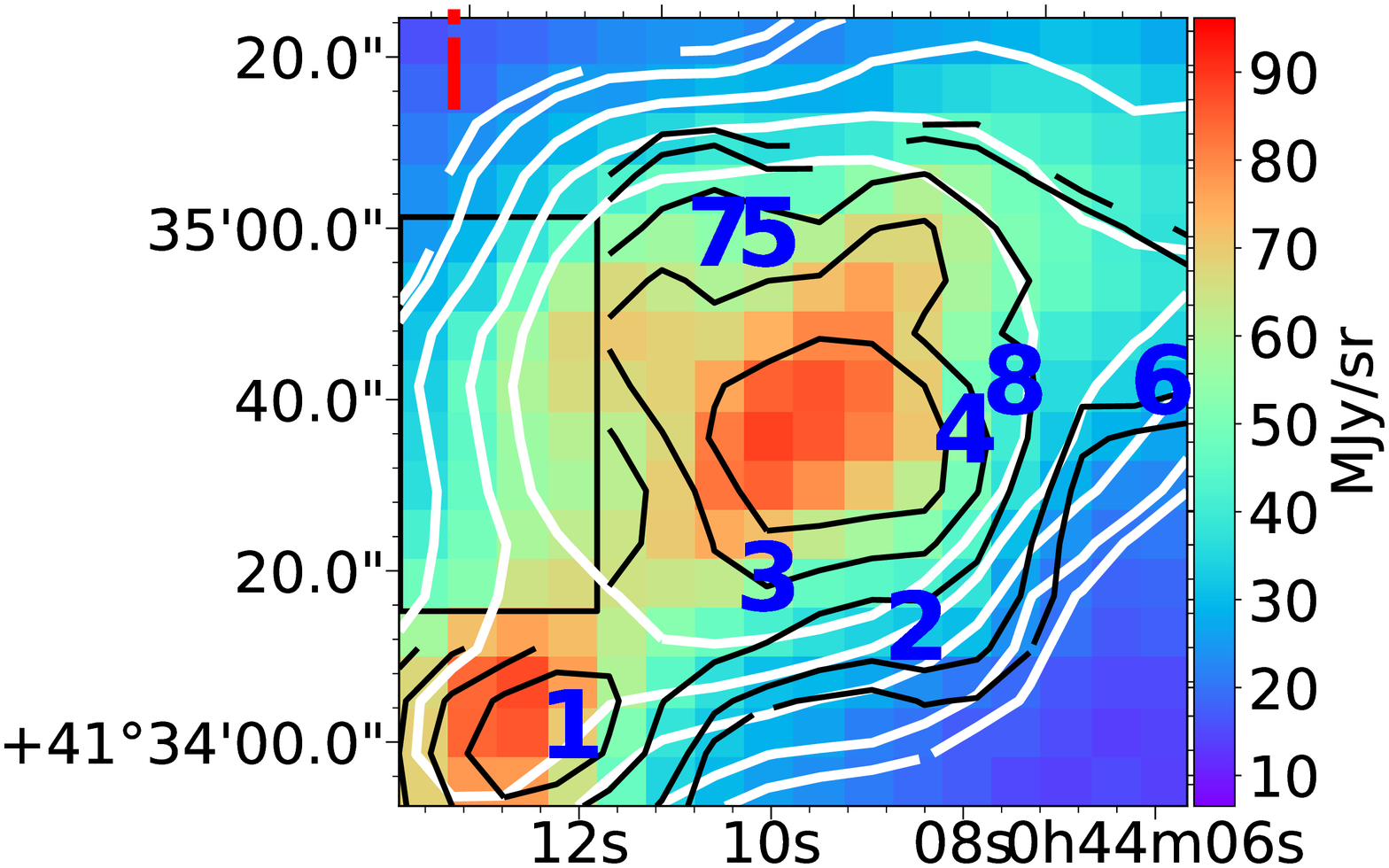}
\includegraphics[width=2.3in]{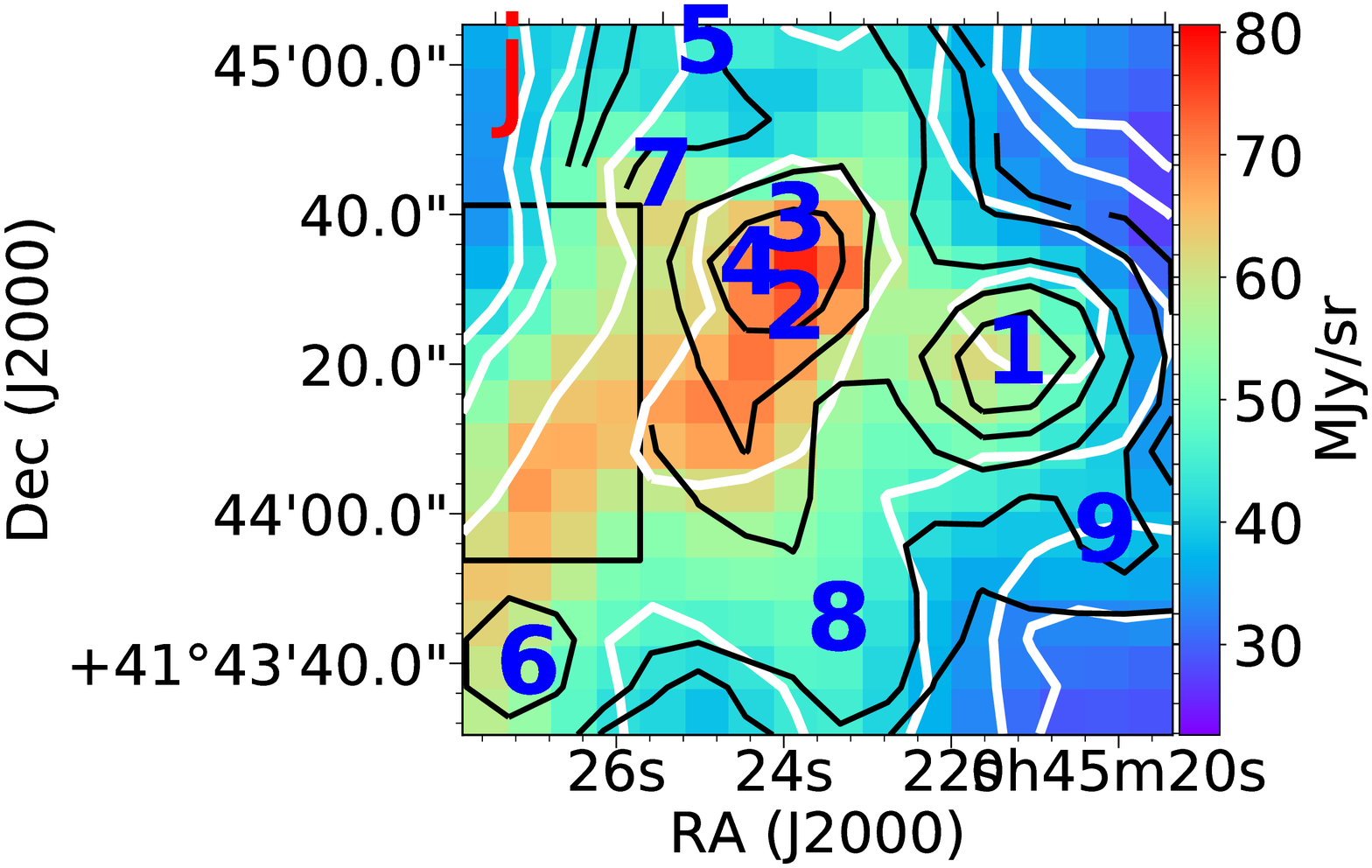}
\includegraphics[width=2.3in]{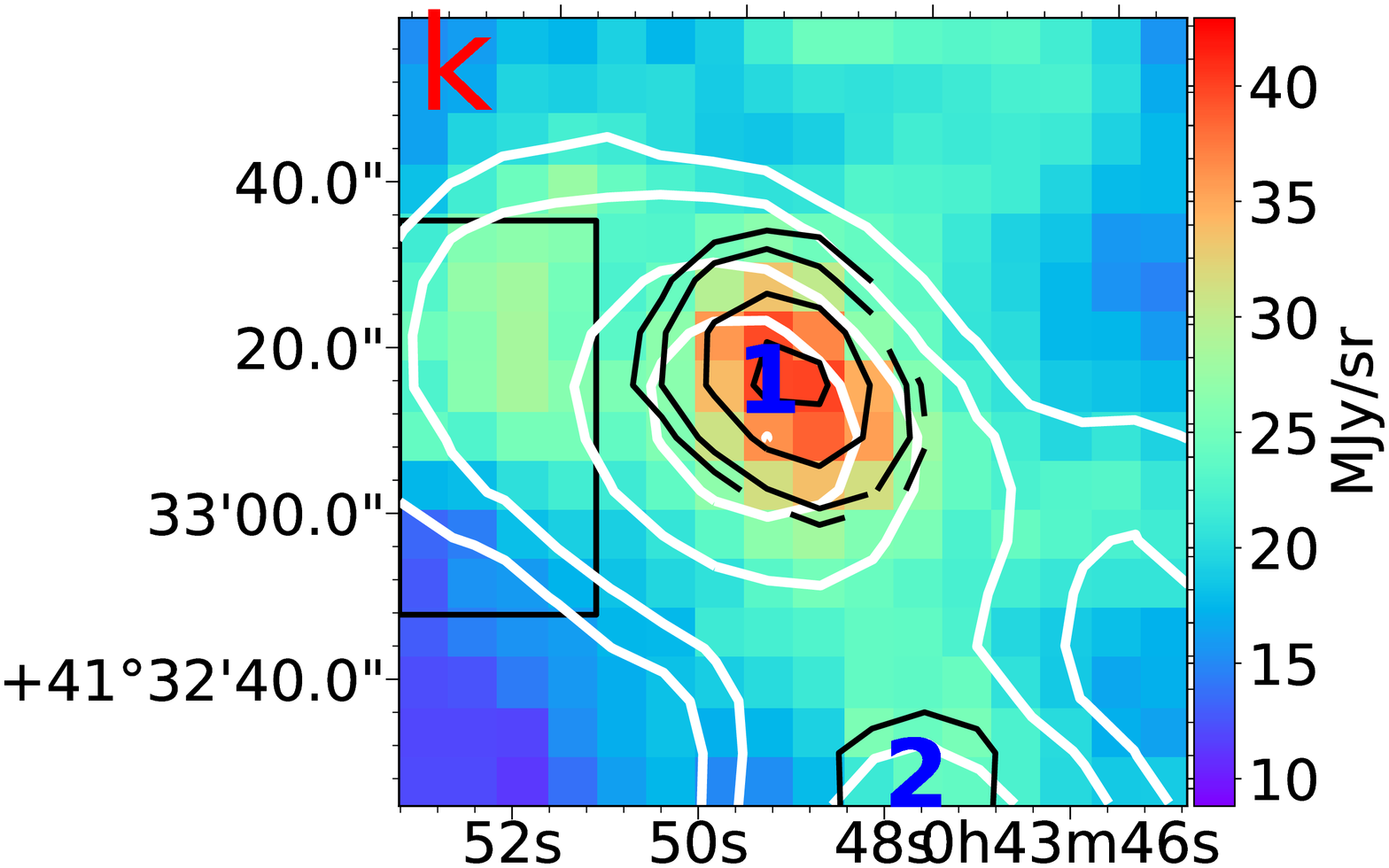}
\caption{Herschel/SPIRE 250 $\mu$m maps of the 11 jiggle fields overlaid with CO(1--0) (white) and CO(3--2) integrated intensity contours (black). The contour levels correspond to 3, 5, 10, 15, 20$\sigma$, respectively, with 1$\sigma$ level of 0.35 K km s$^{-1}$ for CO(1--0) \citep{Nieten}, and 0.10 K km s$^{-1}$ for CO(3--2), which is the typical rms noise in the integrated intensities. 
The physical scale of each field is approximately 500 pc in projection. Field `a' had no CO(1--0) observation. The black rectangles indicate where there are no CO(3--2) observations due to the two receptors that are not operational. The clumps identified by ClumpFind are labelled with numbers. Note that in some cases two or more clumps lie in close proximity along the line-of-sight.\label{fig:2}}
\end{figure}

\begin{figure}
\centering
\includegraphics[width=6in]{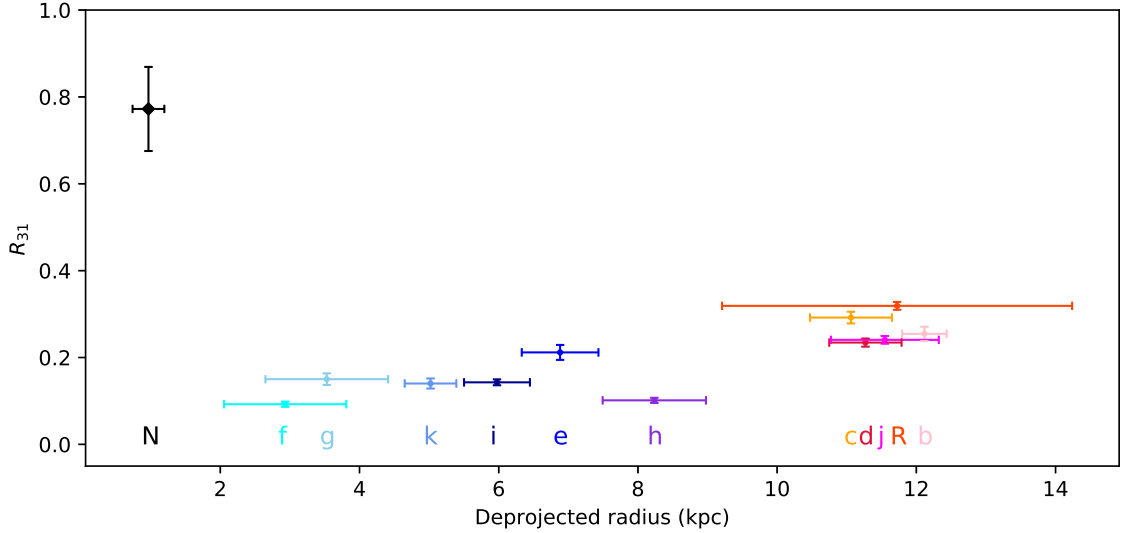}\\
\caption{The radial distribution of CO(3--2)/CO(1--0) intensity ratio, derived from the 11 HASHTAG fields (circles) as well as from a segment of the nuclear ring \citep[diamond, labelled `N', ][]{Li et al. 2019}, same as in Table \ref{tab:2}. The deprojected radius is calculated assuming position angle $38^\circ$ and an inclination angle of $77^\circ$ for the disc \citep{McConnachie 2005}. For the nuclear ring, the inclination angle is assumed to be 45$^\circ$ \citep{Ciardullo 1988}. The vertical and horizontal errorbars represent the uncertainty of the fitted ratio and the radial coverage of each field, respectively.\label{fig:6}}
\end{figure}

There appear to be two branches in Figure \ref{fig:5}. The fields on the 10 kpc ring (with symbols in red-tinted colours) showing a higher CO(3--2) intensity with respect to CO(1--0), while other fields (symbols in blue-tinted colours) exhibit lower values in general. A few points with high line intensities ($I\rm_{CO(3-2)} >$ 3 K km s$^{-1}$ and $I\rm_{CO(1-0)} >$ 6 K km s$^{-1}$), belonging to either JIGGLEd or Raster, are spatially coincident with CO intensity peaks (e.g. clumps d1 and r1), which should trace dense molecular cores that might be the sites of active star formation. 
We applied linear regression to the CO(3--2) and CO(1--0) intensities, using the least-squares fitting method of the $kmpfit$ module\footnote{https://www.astro.rug.nl/software/kapteyn/kmpfit.html} in the Python package $kapteyn$, which accounts for uncertainties in both quantities, to obtain the mean CO(3--2)/CO(1--0) integrated intensity ratio ($R_{31}$). We found that the mean $R_{31}$ values of the five fields on the 10 kpc ring (JIGGLEb, c, d, j and the Raster field) and other fields (JIGGLEe, f, g, h, i, k) are 0.274 $\pm$ 0.005 and 0.137 $\pm$ 0.005, respectively, which are indicated as black lines in Figure \ref{fig:5}. The average line ratio derived from the 11 fields (JIGGLEa has no corresponding CO(1--0) observations) in the disc is 0.228 $\pm$ 0.004. For comparison, the black data points are from the nuclear region \citep{Li et al. 2019}, which shows a substantially higher line ratio of 0.77 $\pm$ 0.10. 
We also calculated the mean $R_{31}$ for each field using the same least-squares fitting method. The results are given in Table \ref{tab:2} and shown as colour-coded dotted lines in Figure \ref{fig:5}. Note that the $R_{31}$ values of the JIGGLEg and h fields are calculated from the `peak' region with CO(3--2) emission, as the CO(1--0) emission is spatially more extended (Figure \ref{fig:2}), hence these values cannot represent the line ratio of the whole field. Combining the measurements of $R_{31}$ in the central kpc obtained by \cite{Li et al. 2019}, Figure \ref{fig:6} illustrates the distribution of $R_{31}$ as a function of the deprojected galactocentric radius. It is clear that the nuclear ring exhibits the highest value of $R_{31}$ ($\sim$0.8), while the 10 kpc ring shows on average a higher line ratio ($\sim$0.27) than the inner disc regions ($\sim$0.14). 
\\

\begin{figure}
\centering
\includegraphics[width=5in]{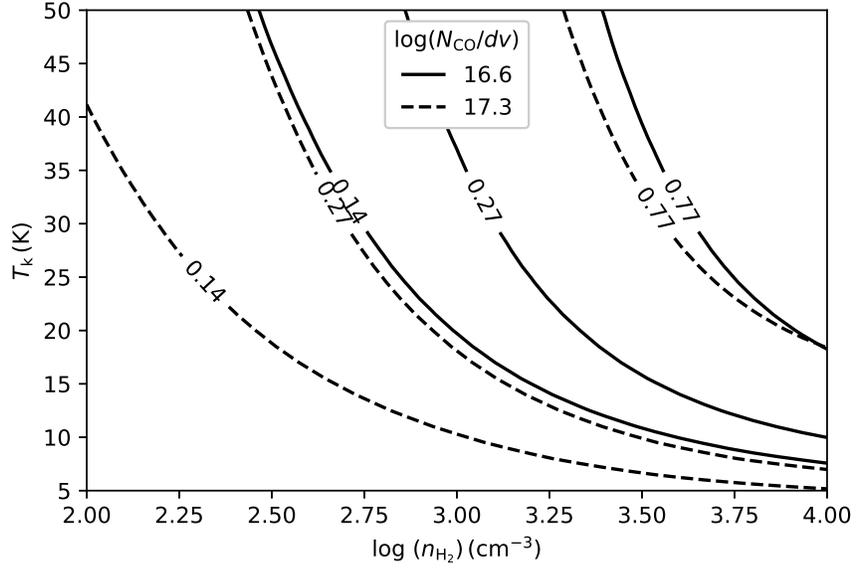}\\
\caption{LVG calculation results for $R_{31}$ = 0.14, 0.27, 0.77, with solid and dashed curves representing log($N\rm_{CO}$/$dv$ (cm$^{-2}$/(km s$^{-1}))$) = 16.6 and 17.3, respectively, which can reproduce the observed line ratios of most HASHTAG fields assuming a CO abundance with respect to H$_2$ of 8 $\times$ 10$^{-5}$. \label{fig:7}}
\end{figure}

We performed radiation transfer calculations of line ratios using the RADEX code \citep{van 2007} to constrain the coupled gas kinetic temperature $T\rm{_k}$ and volume density of molecular hydrogen $n\rm{_{H_2}}$, allowing for optically thick conditions. 
We adopted the large velocity gradient (LVG) approximation model \citep[e.g.][]{Goldreich 1974, Scoville 1974}, which is widely used in the analysis of molecular lines, given the fact that thermal line broadening is far smaller than the velocity dispersion within clouds. The calculations span kinetic temperature
$T\rm{_k}$ = 5 -- 50 K and molecular hydrogen density $n\rm{_{H_2}}$ = 10$^2$ -- 10$^4$ cm$^{-3}$. For the CO column density ($N\rm{_{CO}}$) and velocity gradient ($dv$), we followed \cite{Koda 2012} to use a typical log($N\rm{_{CO}}$/$dv$ (cm$^{-2}$/(km s$^{-1}))$) = 16.6 -- 17.3 based on the observed line widths and assuming a standard ISM CO/H$_2$ abundance of 8$\times$10$^{-5}$ \citep{Schinnerer 2010}. There are a number of caveats pertaining to this analysis that should be kept in mind before interpreting the results. First, the uncertainty in the CO/H$_2$ abundance is not taken into account in the calculations. Moreover, with only two lines, the constraints given by the RADEX code are undetermined and limited. Nevertheless, Figure \ref{fig:7} shows the calculation results for $R_{31}$ = 0.14, 0.27, 0.77, with solid and dashed curves representing log($N\rm_{CO}$/$dv$) = 16.6 and 17.3, respectively. In this plot, the higher $R_{31}$ ratio indicates both higher temperature and higher density in general, with the highest mean ratio of the nuclear region, 0.77, requiring a temperature $\gtrsim$ 20 K and density $\gtrsim$ $10^{3.5}$ cm$^{-3}$, while the lowest value 0.14 only requiring a kinetic temperature $\gtrsim$ 5 K. 
\\

According to the LVG results, the high line ratio in the nuclear region could be due to a high kinetic temperature or a high volume density (or both). However, the low line-of-sight dust extinction toward the M31 centre \citep{Li 2009, Dong et al. 2016} indicates the general absence of very dense and massive molecular clouds. Thus, it is more likely that high temperature is the dominant factor here. This demands heating mechanisms other than star formation or active galactic nucleus (AGN), given the absence of massive stars and nuclear activity in the centre of M31 \citep{Li 2011}. One possibility is heating by old stellar populations that have been claimed to heat the dust in the nuclear region \citep[e.g.][]{Montalto et al. 2009}. Meanwhile, the relatively high line ratio of the five fields on the 10 kpc ring ($\sim$ 0.27) indicates a higher temperature and/or density compared to the inner disc region. This could be due to the difference in star formation activity, since the 10 kpc ring is actively forming stars, while the rest of the disc is quiescent. Among the fields in the inner disc, JIGGLEe has the highest line ratio of 0.21, which may be due to its moderate star formation activity as evidenced by the HII regions therein \citep{Kapala 2015}. 
Two fields (JIGGLEf, g) have been previously suggested to contain very cold gas \citep{Allen 1993, Loinard 1996, Loinard 1998}; although their $R_{31}$ values are indeed low (0.09 and 0.15), these values are comparable to those of other fields in the inner disc (such as JIGGLE h, i, k). 
The fact that the $R_{31}$ ratio in M31's disc ($\sim$0.23) is lower than that of the Milky Way disc \citep[$\sim$0.4,][]{Oka et al. 2012} may suggest that the molecular gas of the M31 disc is on average colder, consistent with the overall low SFR in M31.
\\

\begin{figure}
\centering
\includegraphics[width=5in]{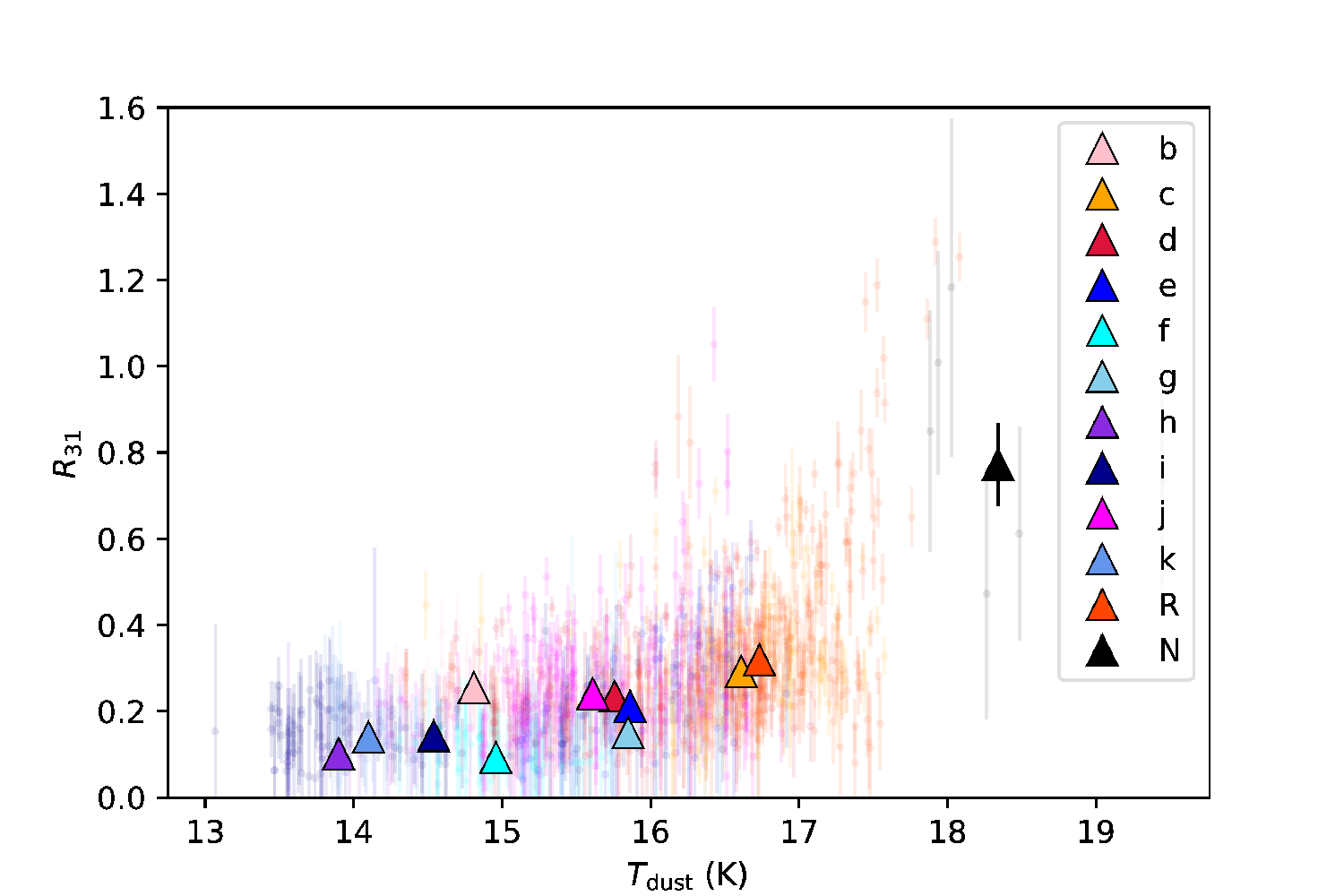}\\
\caption{$R_{31}$ versus dust temperature of each pixel (dots), overlaid with average values of each field (triangles). \label{fig:8}}
\end{figure}

The dust temperature in M31 follows a negative correlation with radius, that is, close to $\sim$30 K in the central region and rapidly decreasing to $\sim$17 K beyond 3 kpc \citep[e.g.][]{Smith 2012, Draine et al. 2014}. In Figure \ref{fig:8}, we contrast $R_{31}$, a plausible indicator of molecular gas temperature, with the dust temperature \citep{Smith 2012} on a pixel-by-pixel basis for a common resolution of 23$\arcsec$. 
To evaluate a possible correlation, we computed the Spearman's rank correlation coefficient $\rho$, which is 0.554 for the whole dataset. The p-value, defined as the probability under the null hypothesis of random variables, is $<$ 0.001. Even after removing data from the nuclear ring, the rank coefficient $\rho$ only decreases a little to 0.546, with p-value $<$ 0.001. The results suggest a positive correlation between $R_{31}$ and dust temperature indeed exists. This correlation could be an implication of the relationship between gas and dust temperature, as $R_{31}$ is a good indicator of gas temperature. 
\\

\begin{deluxetable*}{cccc}[b!]
\tablecaption{Line intensities and line ratios}
\tablecolumns{4}
\tablenum{2}
\tablewidth{2pt}
\tablehead{
\colhead{Field} &
\colhead{$I\rm_{CO(3-2)}$\tablenotemark{(a)}} &
\colhead{$I\rm_{CO(1-0)}$} &
\colhead{$I\rm_{CO(3-2)}$/$I\rm_{CO(1-0)}$\tablenotemark{(b)}} 
}
\startdata
JIGGLEa & 0.50 $\pm$ 0.04 & - & -\\
JIGGLEb & 0.50 $\pm$ 0.03 & 1.89 $\pm$ 0.02 & 0.25 $\pm$ 0.02 \\
JIGGLEc & 1.02 $\pm$ 0.04 & 3.20 $\pm$ 0.08 & 0.29 $\pm$ 0.01 \\
JIGGLEd & 1.32 $\pm$ 0.04 & 4.46 $\pm$ 0.08 & 0.23 $\pm$ 0.01 \\
JIGGLEe & 1.00 $\pm$ 0.04 & 4.19 $\pm$ 0.17 & 0.21 $\pm$ 0.02 \\
JIGGLEf & 0.51 $\pm$ 0.03 & 4.93 $\pm$ 0.15 & 0.09 $\pm$ 0.01 \\
JIGGLEg & 0.19 $\pm$ 0.03 & 1.13 $\pm$ 0.05 & 0.15 $\pm$ 0.01 \\
JIGGLEh & 0.29 $\pm$ 0.02 & 2.82 $\pm$ 0.03 & 0.10 $\pm$ 0.01 \\
JIGGLEi & 0.81 $\pm$ 0.05 & 5.21 $\pm$ 0.10 & 0.14 $\pm$ 0.01 \\
JIGGLEj & 0.84 $\pm$ 0.04 & 3.18 $\pm$ 0.09 & 0.24 $\pm$ 0.01 \\
JIGGLEk & 0.73 $\pm$ 0.04 & 4.60 $\pm$ 0.13 & 0.14 $\pm$ 0.01 \\
Raster & 1.07 $\pm$ 0.04 & 2.95 $\pm$ 0.09 & 0.32 $\pm$ 0.01 \\
\hline
Nucleus\tablenotemark{(c)} & 2.14 $\pm$ 0.22 & 2.38 $\pm$ 0.22 & 0.77 $\pm$ 0.10\\
\enddata
\tablecomments{(a) Mean integrated intensity of CO(3--2) in each field, in units of K km s$^{-1}$, with the statistical uncertainties at 1 $\sigma$ confidence level. (b) Mean ratio and uncertainty calculated using least-squares fitting method. (c) The intensities and line ratio of the nuclear ring. 
\label{tab:2}}
\end{deluxetable*}

\subsection{Correlation with star formation rate}

Previous studies have shown that CO(3--2) emission is more tightly related to star formation than CO(1--0), exhibiting a nearly linear correlation with SFR \citep[e.g.][]{Komugi 2007, Muraoka et al. 2007, Wilson 2009}. Similarly, \cite{Wilson 2012} compared the CO(3--2) luminosity of the NGLS sample with the far-infrared luminosity from the $Spitzer$ Infrared Nearby Galaxies Survey (SINGS), finding that they are nearly linearly correlated. As for M31, \cite{Leroy 2008}, \cite{Ford 2013} and \cite{Rahmani 2016} have claimed that the relationship between the SFR surface density ($\Sigma\rm_{SFR}$) and the gas column density ($\Sigma\rm_{mol}$; i.e. the KS law) varies radially in M31, the first time that such a radial variation was reported for a galaxy. Furthermore, they found a sub-linear relationship between $\Sigma\rm_{SFR}$ and $\Sigma\rm_{mol}$ using the CO(1--0) survey data, which indicates the presence of CO-bright regions that are not efficiently forming stars.
\\

\begin{figure}
\centering
\includegraphics[width=3.5in]{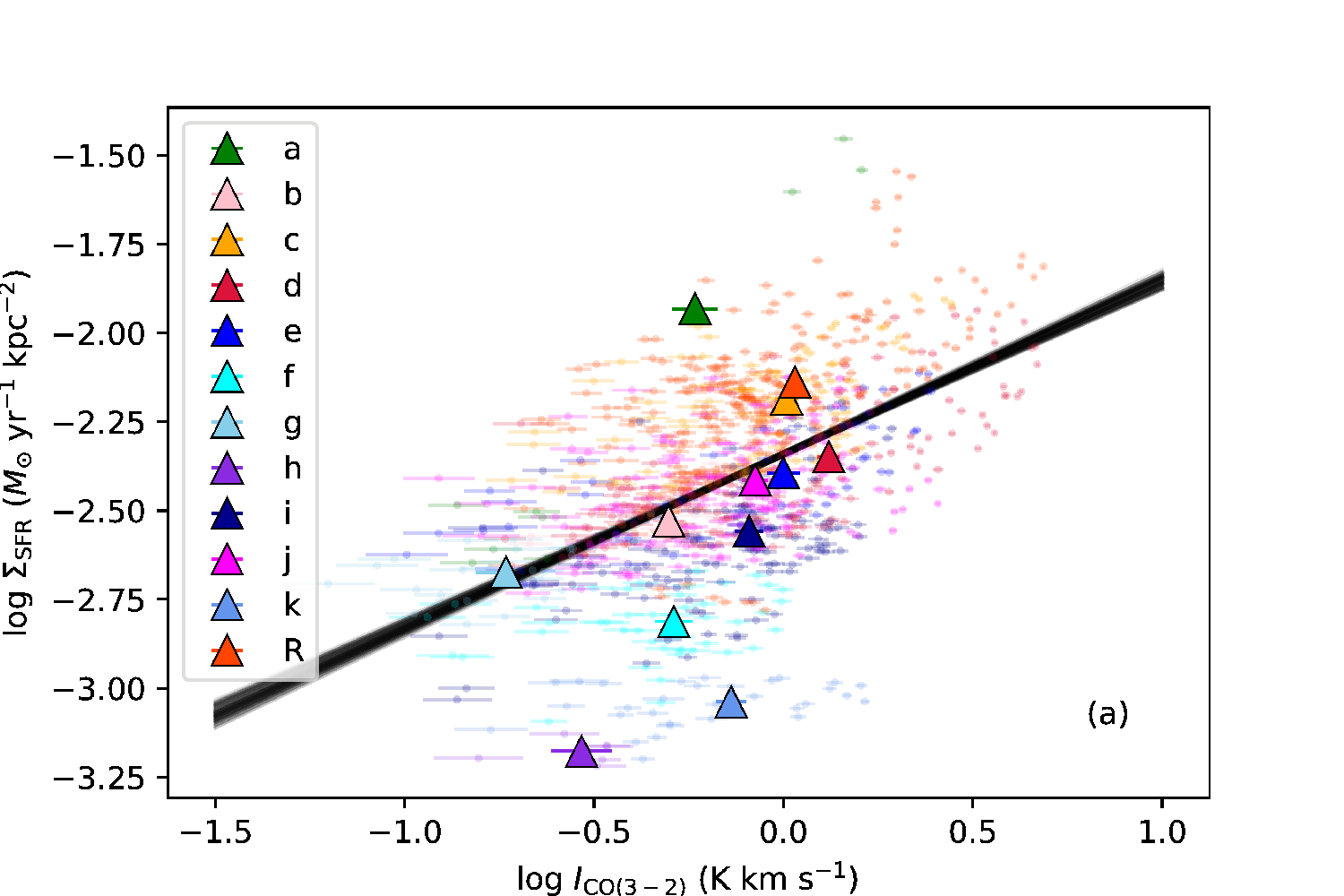}
\includegraphics[width=3.5in]{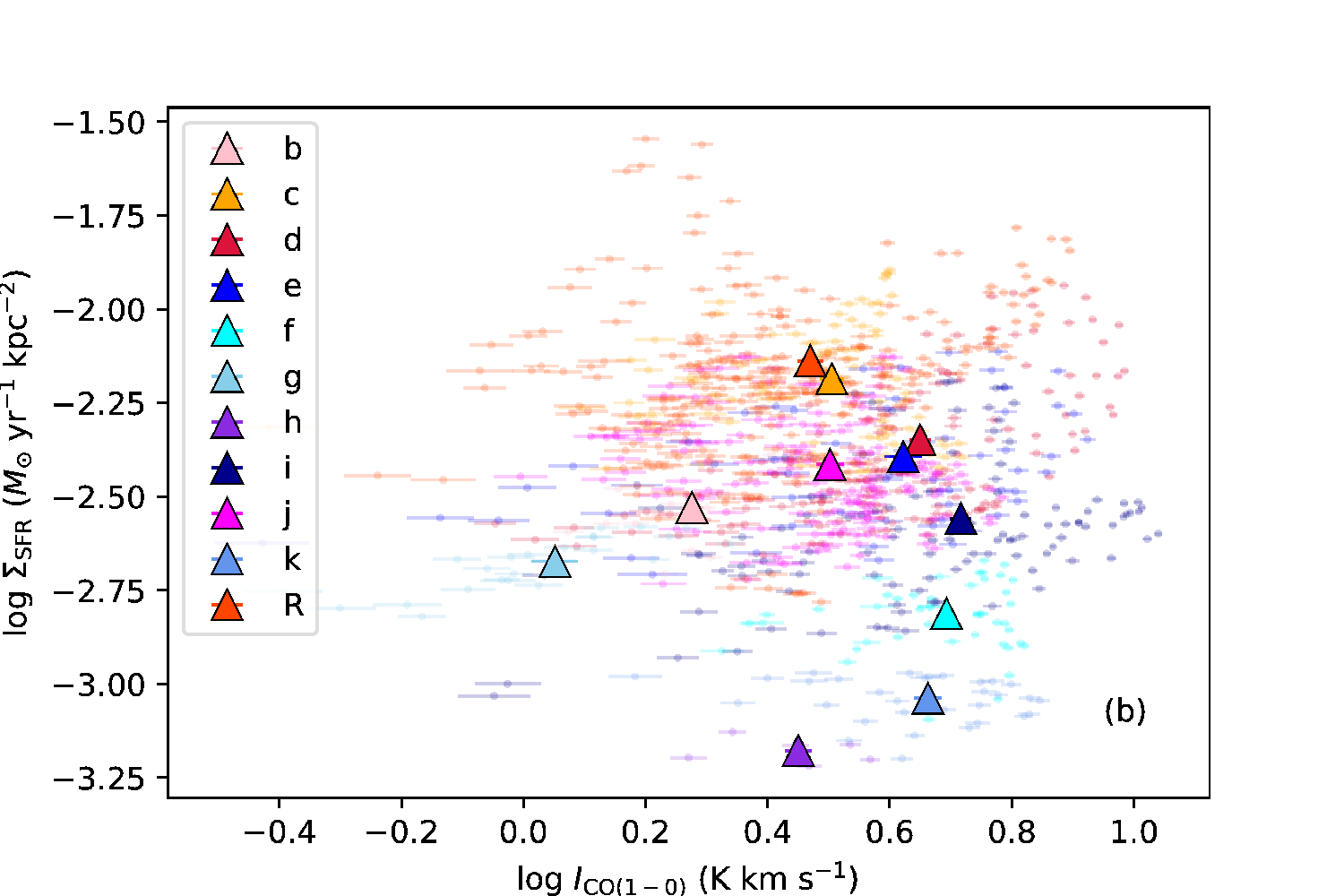}
\includegraphics[width=3.5in]{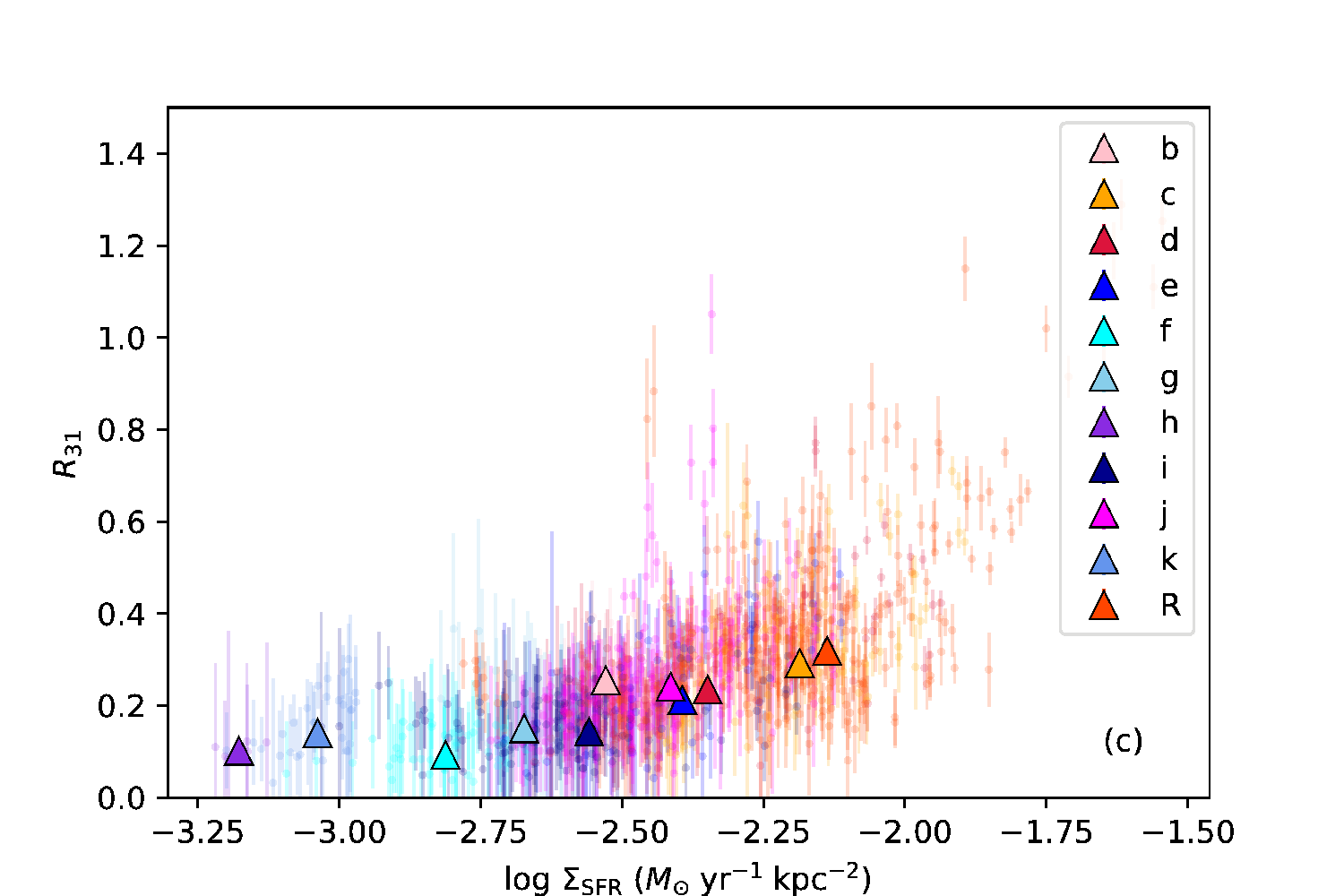}\\
\caption{(a) Pixel-by-pixel (small dots) correlation of SFR surface density and CO(3--2) integrated intensity. The triangles represent the mean values of each field. The black strip represent a best-fitting power-law index of $0.49^{+0.03}_{-0.03}$. (b) Comparison of SFR surface density and CO(1--0) integrated intensity, with symbols the same as in (a). (c) Comparison of $R_{31}$ ratio and SFR surface density $\Sigma\rm_{SFR}$ of each pixel (dot) and field (triangle). 
\label{fig:9}}
\end{figure}

We examined the relationship between the CO emission and the SFR surface density of \cite{Ford 2013}. The latter was convolved to a common resolution of 23$\arcsec$, with a typical uncertainty of 2.4 $\times$ 10$^{-4}$ $M_{\odot} \rm \; yr^{-1} \; kpc^{-2}$. 
Since the commonly used conversion factor $\alpha\rm{_{CO}}$=$\Sigma\rm_{mol}/\it I\rm{_{CO(1-0)}}$ \citep{Bolatto et al. 2013} is of substantial uncertainty \citep{Clark 2015, Shetty 2011}, and there is no single factor for CO(3--2) to apply, we investigated the relationship among SFR surface density, $\Sigma\rm_{SFR}$, and CO integrated intensity, $I\rm_{CO}$, instead of the molecular gas surface density. 
As shown in Figure \ref{fig:9}(a), the relationship is examined pixel-by-pixel (small dots), overlaid with the mean values of each field (triangles) on a characteristic scale of $\sim$500 pc. 
\\

The correlation between the SFR surface density and CO(3--2) integrated intensity for the 12 HASHTAG fields is relatively strong, with Spearman's rank correlation coefficient $\rho$ = 0.54 (p $<$ 0.001). We thus fitted a linear relationship between them on a pixel-by-pixel basis, using the Markov chain Monte Carlo (MCMC) method implemented in the Python package $emcee$ \citep{Foreman-Mackey 2013}, following \cite{Sun 2018}. We found a best-fitting power-law index $0.49^{+0.03}_{-0.03}$, as shown by the black strip in Figure \ref{fig:9}(a). This index is roughly consistent with that found by \cite{Rahmani 2016} in their analysis of H$_2$-only gas, based on CO(1--0) conversion, which is 0.540 $\pm$ 0.003 for the whole galaxy. However, it is still lower than that reported by both \cite{Tabatabaei 2010} ($\sim$1.0) and \cite{Ford 2013} ($\sim$0.6). 
We note that the JIGGLE h and k fields have much lower SFR surface densities than the other fields ($< 10^{-3}\; M_{\odot}\; \rm yr^{-1}\; kpc^{-2}$); both lie in the interarm region where little star formation is observed. 
If we were to exclude these two fields as outliers, the power-law index will become even lower, 0.46 $\pm$ 0.02. Therefore, the selection criteria of data points, such as the choice of outliers and an S/N cut \citep{Williams 2018}, may affect the power-law index. 
For comparison, \cite{Williams 2018} found a super-linear correlation between $\Sigma\rm_{SFR}$ and $\Sigma\rm_{H_2}$ in M33 on a similar sub-kpc scale. This suggests that the KS law may vary substantially from one galaxy to another, as claimed by \cite{Shetty 2013}. 
\\

On the other hand, the CO(1--0) emission from the HASHTAG fields shows no significant correlation with the SFR surface density, with rank coefficient $\rho$ = --0.05 (p = 0.16). This may be explained by our limited number of data points compared with other studies, and the relation we discussed here is all based on small regions with characteristic sizes $\lesssim$500 pc, where the relation may not hold \citep{Tabatabaei 2010, Ford 2013, Querejeta 2019}. On small scales, other factors than the molecular gas surface density may play a key role in determining the star formation activity \citep{Bigiel et al. 2008, Leroy 2008}, such as gas turbulence \citep{Sun 2018} and stellar mass surface density \citep{Shi 2018}. 
\\

We also examined the correlation between $R_{31}$ and the SFR surface density, as shown in Figure \ref{fig:9}(c). 
The Spearman's coefficient for this correlation is 0.69 (p $<$ 0.001), suggesting a significant correlation between SFR and $R_{31}$. Interestingly, \cite{Viaene 2018} found a significant correlation between the SFR and HCN line emission, a commonly used tracer of dense gas \citep{Gao 2004}, in a few individual giant molecular clouds of M31; they found an even tighter correlation between the SFR and the HCN/CO(1--0) ratio, which led them to propose a dependency of the SFR on the dense gas or dense gas fraction. Since CO(3--2) is sensitive to warm and dense gas, the relatively tight correlations between the SFR and the CO(3--2) intensity as well as the CO(3--2)/CO(1--0) ratio may suggest a dependency of the SFR on the warm gas density or warm gas fraction as well. 
It is noteworthy that almost all the points in the upper right part of Figure \ref{fig:9}c are from the 10 kpc star forming ring (the JIGGLEc, j and Raster fields), with higher line ratio, higher SFR surface density, and also, higher dust temperature. Thus, we suggest that the systematically higher values of $R_{31}$ in the 10 kpc ring, compared to the inner disc, are at least partly due to star formation heating. Indeed, according to \cite{Viaene et al. 2017}, the overall dust heating in M31 is dominated by evolved (i.e., old) stellar populations. However, the relative contributions of young stellar populations (i.e., younger than 100 Myr) in certain regions along the 10 kpc ring can be as high as 20\%--50\%, compared with the global fraction of 10\%.
\\

\section{Summary}

We have presented a systematic survey of CO(3--2) emission lines from 12 selected regions across the M31 disc, observed as part of the HASHTAG programme. 
The data will be publicly available on the HASHTAG website\footnote{ https://www.eao.hawaii.edu/HASHTAG }. We use these data as well as existing CO(3--2) and CO(1--0) observations and complementary multi-wavelength images to investigate the properties of the molecular gas and its relation with dust and star formation activity in regions spanning a galactocentric radial range of 1--14 kpc. The main results are as follows:

\begin{itemize}

\item Except for JIGGLEa which lies at the far side of the disc and was not covered by the CO(1--0) survey, the CO(3--2) lines of all other 11 fields are much weaker than the CO(1--0) lines. 
We found that the $R_{31}$ ratio of the M31 disc slightly increases with radius, from $\sim 0.14$ within 10 kpc to $\sim$ 0.27 in the 10 kpc ring, while the ratio of the nuclear ring (at $\sim$1 kpc), $\sim$0.8, is significantly higher. 

\item We conducted LVG calculations to reproduce the line ratios in the $n\rm_{H_2}$ and $T\rm_k$ plane, which requires higher kinetic temperature and volume density for higher line ratios in general. 
In the star forming 10 kpc ring, the higher line ratio may be explained by heating induced by star formation activity. Nevertheless, the high ratio and implied high temperature of the nuclear ring needs other heating mechanisms given the lack of nuclear star formation and AGN activity in M31.
We also found a weak positive correlation between the dust temperature and the $R_{31}$ ratio. 

\item We also investigated the relation between $\Sigma\rm_{SFR}$ and $I\rm_{CO}$ of CO(3--2) and CO(1--0), respectively. We found a tighter relationship between the SFR surface density and CO(3--2) integrated intensity than with CO(1--0), consistent with previous studies. We also found a significant correlation with $R_{31}$ and $\Sigma\rm_{SFR}$.

\end{itemize}

In this paper, we have focused on presenting the results of the CO(3--2) observations, and analyses of the $R_{31}$ ratio. More in-depth investigation of the properties of cold ISM across the M31 disc, combining HASHTAG observations and complementary multi-wavelength data, including $Herschel$, PHAT imaging and other CO transitions, will be the subject of future work. 

\begin{acknowledgements}

The James Clerk Maxwell Telescope is operated by the East Asian Observatory on behalf of The National Astronomical Observatory of Japan; Academia Sinica Institute of Astronomy and Astrophysics; the Korea Astronomy and Space Science Institute; Center for Astronomical Mega-Science (as well as the National Key R\&D Program of China with No. 2017YFA0402700). Additional funding support is provided by the Science and Technology Facilities Council of the United Kingdom and participating universities in the United Kingdom and Canada. Data for the JINGLE Large Program have been obtained under the JCMT project code M17BL005. Z.N.L., Z.Y.L. and Y.G. acknowledge support by the National Key Research and Development Program of China (2017YFA0402703) and National Natural Science Foundation of China (grant 11873028). C.D.W. acknowledges support from the Natural Science and Engineering Research Council of Canada and the Canada Research Chairs program. T.M.H. acknowledges support from the Chinese Academy of Sciences (CAS) and the National Commission for Scientific and Technological Research of Chile (CONICYT) through a CAS-CONICYT Joint Postdoctoral Fellowship administered by the CAS South America Center for Astronomy (CASSACA) and CONICYT in Santiago, Chile. F.K. and P.S. acknowledge support from the Ministry of Science and Technology of Taiwan under grant MOST107-2119-M-001-031-MY3 and from Academia Sinica under grant AS-IA-106-M03. M.J.M. acknowledges the support of  the National Science Centre, Poland through the SONATA BIS grant 2018/30/E/ST9/00208. J.H. thanks the National Natural Science Foundation of China under Grant Nos. 11873086 and U1631237 and the support by Yunnan Province of China (No.2017HC018). This work is sponsored (in part) by the Chinese Academy of Sciences (CAS), through a grant to the CAS South America Center for Astronomy (CASSACA) in Santiago, Chile.

\end{acknowledgements}

\clearpage

\appendix

The pixel-by-pixel comparison of the line-of-sight velocity and velocity dispersion between the CO(3--2) and CO(1--0) lines, after applying the criteria described in Section 3.2, is shown in Figure \ref{fig:a1}. Different velocity components of each field are evident in the top panel, and it is clear that the two lines have a consistent velocity dispersion despite the large scatter.
\\

We list the 69 clumps found with the ClumpFind algorithm in $findclump$ task in Table \ref{tab:3}. They are listed in order of the field name. For example, clump ``a1" is the first clump found in JIGGLEa field. Their positions, sizes, peak and integrated intensities are given in the table. 
The clumps have a typical size (FWHM) of $\sim$90 pc, and a typical line width of $\sim$9 km s$^{-1}$.
\\

\setcounter{figure}{0}
\renewcommand\thefigure{A\arabic{figure}}
\begin{figure}
\centering
\includegraphics[width=5in]{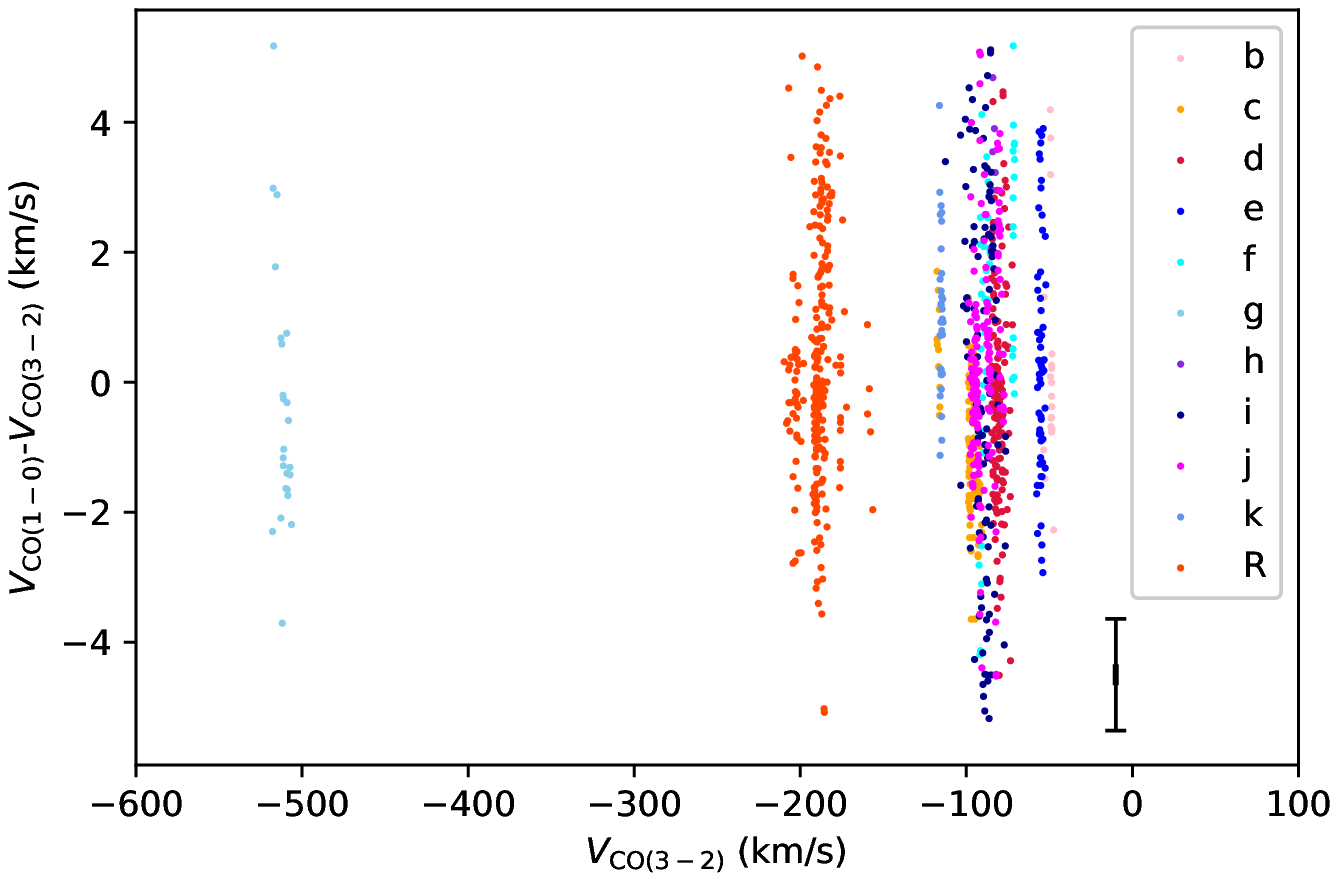}
\includegraphics[width=5in]{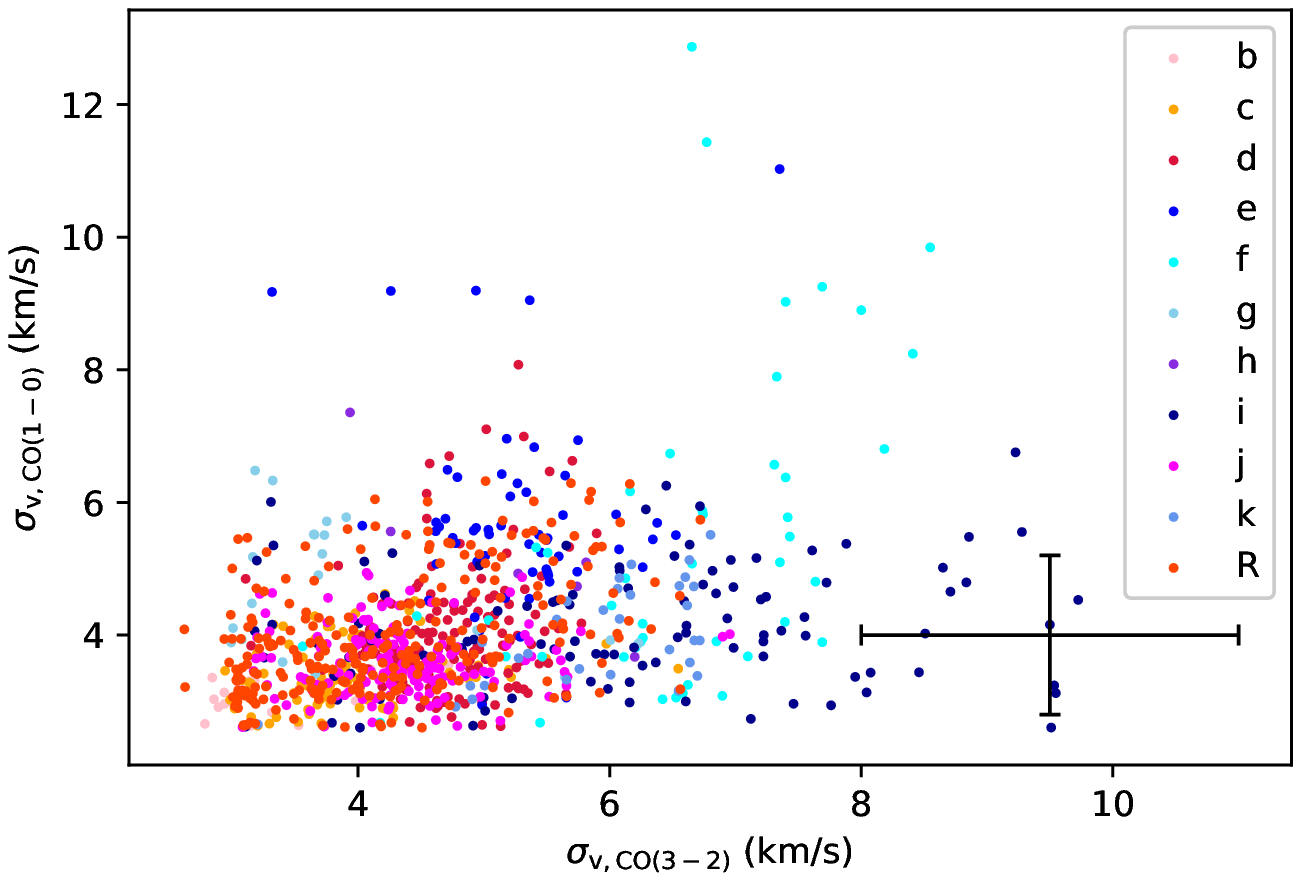}
\caption{Pixel-by-pixel comparison of the line-of-sight velocity (top panel) and velocity dispersion (bottom panel) between CO(3--2) and CO(1--0) in all HASHTAG fields. Each pixel is obtained from the moment maps based on individual clumps found by ClumpFind. We only consider the two lines to arise from the same component if the difference of their central velocity is less than two channel widths, 5.2 km s$^{-1}$. We also eliminated pixels with velocity dispersion narrower than the channel width 2.6 km s$^{-1}$. The typical uncertainty is shown as black cross at the lower right. \label{fig:a1}}
\end{figure}

\begin{deluxetable*}{ccccccccc}
\tablecaption{Catalog of Identified Molecular Clumps}
\tablenum{3}
\tablewidth{0pt}
\tablehead{
\colhead{ID} &\colhead{RA (J2000)}&\colhead{Dec (J2000)}&
\colhead{$v$} &\colhead{$D_x$} &\colhead{$D_y$}&\colhead{$D_{v}$}
&\colhead{$T\rm_{peak}$}&\colhead{$I\rm_{int}$}\\
\colhead{} &\colhead{($^\circ$)}&\colhead{($^\circ$)}&
\colhead{(km s$^{-1}$)} &\colhead{(pc)} &\colhead{(pc)}&\colhead{(km s$^{-1}$)}
&\colhead{(K)}&\colhead{(K km s$^{-1}$)}
}
\decimalcolnumbers
\startdata
a1 & 11.6400 & 42.1933 & --45.1 & 108.9 & 90.1 & 5.8 & 0.23 & 0.49 $\pm$ 0.05\\
a2 & 11.6203 & 42.1892 & --44.9 & 111.4 & 163.1 & 5.2 & 0.06 & 0.12 $\pm$ 0.02\\
b1 & 11.3861 & 41.9867 & --53.3 & 102.7 & 64.1 & 5.7 & 0.11 & 0.25 $\pm$ 0.03\\
b2 & 11.3880 & 41.9841 & --45.3 & 87.3 & 74.7 & 5.5 & 0.12 & 0.28 $\pm$ 0.03\\
b3 & 11.4001 & 41.9721 & --49.5 & 84.9 & 83.0 & 6.4 & 0.09 & 0.34 $\pm$ 0.04\\
b4 & 11.3892 & 41.9727 & --48.9 & 84.5 & 88.5 & 5.1 & 0.08 & 0.25 $\pm$ 0.03\\
c1 & 11.1425 & 41.8794 & --96.6 & 119.1 & 102.2 & 7.3 & 0.31 & 0.87 $\pm$ 0.04\\
c2 & 11.1314 & 41.8772 & --95.0 & 57.3 & 115.2 & 5.6 & 0.23 & 0.80 $\pm$ 0.05\\
c3 & 11.1637 & 41.8827 & --99.1 & 147.4 & 108.4 & 7.8 & 0.19 & 0.59 $\pm$ 0.05\\
c4 & 11.1779 & 41.8738 & --94.0 & 72.6 & 86.5 & 6.7 & 0.17 & 0.64 $\pm$ 0.04\\
c5 & 11.1493 & 41.8607 & --116.5 & 73.8 & 71.7 & 6.5 & 0.13 & 0.48 $\pm$ 0.04\\
c6 & 11.1461 & 41.8776 & --88.4 & 86.3 & 81.6 & 6.7 & 0.07 & 0.20 $\pm$ 0.03\\
c7 & 11.1548 & 41.8851 & --110.4 & 88.1 & 93.3 & 7.1 & 0.11 & 0.29 $\pm$ 0.04\\
d1 & 11.2350 & 41.9257 & --81.2 & 139.9 & 107.9 & 9.3 & 0.49 & 1.88 $\pm$ 0.05\\
d2 & 11.2471 & 41.9193 & --83.5 & 119.2 & 85.8 & 8.7 & 0.23 & 0.63 $\pm$ 0.04\\
d3 & 11.2653 & 41.9189 & --79.5 & 106.7 & 109.5 & 7.6 & 0.14 & 0.62 $\pm$ 0.04\\
d4 & 11.2604 & 41.9240 & --74.7 & 100.5 & 101.6 & 7.6 & 0.16 & 0.45 $\pm$ 0.03\\
d5 & 11.2266 & 41.9176 & --81.8 & 78.6 & 152.9 & 11.6 & 0.09 & 0.34 $\pm$ 0.04\\
d6 & 11.2541 & 41.9306 & --71.3 & 104.7 & 125.5 & 7.3 & 0.07 & 0.19 $\pm$ 0.03\\
e1 & 11.1074 & 41.6287 & --55.5 & 136.4 & 111.4 & 12.8 & 0.22 & 0.75 $\pm$ 0.04\\
f1 & 10.7719 & 41.4064 & --88.1 & 118.2 & 87.4 & 15.1 & 0.10 & 0.42 $\pm$ 0.04\\
f2 & 10.7804 & 41.4099 & --68.6 & 122.6 & 81.1 & 8.5 & 0.08 & 0.31 $\pm$ 0.04\\
f3 & 10.7631 & 41.4089 & --88.2 & 89.8 & 68.8 & 19.9 & 0.05 & 0.22 $\pm$ 0.03\\
g1 & 10.5999 & 41.1103 & --514.1 & 92.2 & 129.4 & 12.5 & 0.04 & 0.13 $\pm$ 0.03\\
g2 & 10.5875 & 41.0921 & --508.7 & 107.4 & 116.9 & 7.0 & 0.04 & 0.13 $\pm$ 0.03\\
g3 & 10.5860 & 41.1143 & --576.5 & 157.7 & 145.8 & 5.5 & 0.02 & 0.04 $\pm$ 0.02\\
h1 & 11.0156 & 41.7101 & --81.2 & 118.1 & 139.0 & 10.1 & 0.05 & 0.13 $\pm$ 0.03\\
i1 & 11.0545 & 41.5728 & --82.8 & 101.7 & 95.1 & 9.0 & 0.26 & 0.78 $\pm$ 0.04\\
i2 & 11.0425 & 41.5856 & --95.2 & 75.5 & 84.0 & 6.9 & 0.10 & 0.37 $\pm$ 0.03\\
i3 & 11.0536 & 41.5835 & --91.5 & 93.7 & 121.0 & 12.0 & 0.13 & 0.71 $\pm$ 0.05\\
i4 & 11.0514 & 41.5893 & --79.0 & 129.4 & 90.0 & 9.9 & 0.11 & 0.54 $\pm$ 0.04\\
i5 & 11.0607 & 41.5924 & --93.1 & 142.3 & 92.8 & 10.9 & 0.12 & 0.59 $\pm$ 0.05\\
i6 & 11.0448 & 41.5989 & --98.0 & 74.4 & 94.0 & 9.0 & 0.08 & 0.38 $\pm$ 0.04\\
i7 & 11.0631 & 41.5901 & --104.0 & 86.2 & 105.1 & 10.2 & 0.06 & 0.24 $\pm$ 0.04\\
i8 & 11.0490 & 41.5948 & --90.1 & 85.0 & 121.1 & 7.9 & 0.10 & 0.32 $\pm$ 0.04\\
j1 & 11.3529 & 41.7552 & --86.0 & 103.0 & 101.8 & 7.1 & 0.32 & 0.72 $\pm$ 0.04\\
j2 & 11.3650 & 41.7479 & --93.6 & 98.6 & 91.9 & 8.4 & 0.18 & 0.78 $\pm$ 0.05\\
j3 & 11.3694 & 41.7574 & --94.4 & 94.8 & 89.2 & 7.3 & 0.19 & 0.68 $\pm$ 0.04\\
j4 & 11.3705 & 41.7505 & --78.0 & 87.5 & 105.7 & 7.8 & 0.15 & 0.45 $\pm$ 0.03\\
j5 & 11.3773 & 41.7580 & --87.3 & 117.3 & 40.5 & 9.2 & 0.10 & 0.50 $\pm$ 0.03\\
j6 & 11.3651 & 41.7314 & --96.9 & 86.6 & 73.8 & 6.9 & 0.17 & 0.74 $\pm$ 0.05\\
j7 & 11.3763 & 41.7511 & --86.6 & 78.4 & 89.1 & 7.5 & 0.08 & 0.31 $\pm$ 0.03\\
j8 & 11.3538 & 41.7411 & --95.9 & 115.1 & 104.7 & 8.7 & 0.11 & 0.50 $\pm$ 0.04\\
j9 & 11.3438 & 41.7504 & --92.2 & 93.0 & 76.6 & 5.8 & 0.09 & 0.30 $\pm$ 0.03\\
k1 & 10.9698 & 41.5649 & --113.9 & 88.9 & 114.3 & 9.7 & 0.13 & 0.45 $\pm$ 0.04\\
k2 & 10.9511 & 41.5558 & --115.0 & 68.4 & 61.3 & 6.6 & 0.06 & 0.26 $\pm$ 0.03\\
\enddata
\end{deluxetable*}

\begin{deluxetable*}{ccccccccc}
\tablecaption{Catalog of Identified Molecular Clumps}
\tablecolumns{12}
\tablenum{3}
\tablewidth{0pt}
\tablehead{
\colhead{ID} &\colhead{RA (J2000)}&\colhead{Dec (J2000)}&
\colhead{$v$} &\colhead{$D_x$} &\colhead{$D_y$}&\colhead{$D_{v}$}
&\colhead{$T\rm_{peak}$}&\colhead{$I\rm_{int}$}\\
\colhead{} &\colhead{($^\circ$)}&\colhead{($^\circ$)}&
\colhead{(km s$^{-1}$)} &\colhead{(pc)} &\colhead{(pc)}&\colhead{(km s$^{-1}$)}
&\colhead{(K)}&\colhead{(K km s$^{-1}$)}
}
\decimalcolnumbers
\startdata
R1 & 11.1832 & 41.4663 & --185.5 & 124.5 & 109.3 & 10.1 & 0.45 & 1.42 $\pm$ 0.04\\
R2 & 11.1913 & 41.4590 & --191.3 & 115.1 & 87.2 & 11.3 & 0.21 & 1.07 $\pm$ 0.04\\
R3 & 11.1584 & 41.4205 & --202.3 & 138.9 & 103.9 & 10.1 & 0.22 & 0.78 $\pm$ 0.04\\
R4 & 11.1702 & 41.4218 & --202.9 & 108.0 & 86.5 & 11.8 & 0.15 & 0.58 $\pm$ 0.03\\
R5 & 11.1787 & 41.4345 & --190.7 & 86.4 & 82.0 & 6.7 & 0.14 & 0.52 $\pm$ 0.03\\
R6 & 11.2097 & 41.4927 & --158.7 & 75.2 & 53.7 & 6.9 & 0.21 & 0.83 $\pm$ 0.06\\
R7 & 11.1953 & 41.4452 & --213.1 & 99.7 & 76.1 & 4.4 & 0.15 & 0.34 $\pm$ 0.03\\
R8 & 11.1714 & 41.4401 & --189.3 & 75.5 & 78.7 & 6.5 & 0.11 & 0.42 $\pm$ 0.03\\
R9 & 11.1700 & 41.4489 & --191.0 & 90.2 & 84.3 & 7.5 & 0.11 & 0.56 $\pm$ 0.03\\
R10 & 11.1413 & 41.4847 & --175.9 & 139.6 & 82.0 & 5.7 & 0.11 & 0.34 $\pm$ 0.02\\
R11 & 11.1354 & 41.4262 & --206.6 & 88.8 & 103.3 & 7.6 & 0.15 & 0.49 $\pm$ 0.03\\
R12 & 11.2016 & 41.4834 & --157.6 & 114.6 & 72.7 & 5.0 & 0.11 & 0.41 $\pm$ 0.03\\
R13 & 11.2099 & 41.4646 & --179.1 & 97.5 & 89.3 & 6.7 & 0.12 & 0.32 $\pm$ 0.03\\
R14 & 11.1836 & 41.4521 & --188.9 & 81.4 & 94.3 & 9.3 & 0.11 & 0.57 $\pm$ 0.03\\
R15 & 11.1622 & 41.4549 & --192.0 & 107.7 & 73.7 & 7.5 & 0.11 & 0.53 $\pm$ 0.03\\
R16 & 11.1720 & 41.4521 & --183.5 & 81.7 & 85.7 & 8.4 & 0.09 & 0.46 $\pm$ 0.03\\
R17 & 11.1637 & 41.4648 & --189.4 & 95.6 & 105.9 & 4.7 & 0.07 & 0.25 $\pm$ 0.02\\
R18 & 11.2165 & 41.4703 & --172.4 & 69.3 & 66.3 & 6.7 & 0.08 & 0.29 $\pm$ 0.02\\
R19 & 11.2234 & 41.4758 & --174.9 & 93.6 & 83.2 & 6.0 & 0.13 & 0.37 $\pm$ 0.03\\
R20 & 11.2006 & 41.4731 & --181.3 & 106.6 & 108.6 & 8.9 & 0.10 & 0.42 $\pm$ 0.03\\
R21 & 11.2398 & 41.4753 & --202.1 & 159.1 & 125.9 & 3.7 & 0.36 & 0.55 $\pm$ 0.07\\
R22 & 11.1873 & 41.4792 & --165.1 & 88.7 & 68.4 & 9.1 & 0.09 & 0.33 $\pm$ 0.02\\
R23 & 11.1778 & 41.4788 & --181.1 & 109.5 & 94.7 & 9.4 & 0.07 & 0.30 $\pm$ 0.03\\
\enddata
\tablecomments{(1) The ID of molecular clumps within each field. (2)--(4) 
Coordinates of the centroid of each clump. (5)--(7) The clump size (FWHM) in three dimensions, calculated using the RMS deviation of each pixel centre from the clump centroid, where each pixel is weighted by the corresponding pixel data value, i.e. CO(3--2) brightness temperature. 
(8) Peak brightness temperature of the clump. The RMS of this value is similar to the typical RMS of each field listed in Table \ref{tab:1}. (9) Integrated intensity of the clump $\pm$ measurement uncertainty. 
\label{tab:3}}
\end{deluxetable*}


\clearpage

\begin{thebibliography}{}

\bibitem[Allen \& Lequeux(1993)]{Allen 1993}Allen, R. J., \& Lequeux, J. \ 1993, ApJ, 410, L15
\bibitem[Banerji et al.(2009)]{Banerji 2009}Banerji, M., Viti, S., \& Williams, D. A. 2009, ApJ, 703, 2249
\bibitem[Beaton et al.(2007)]{Beaton 2007}Beaton, R. L., Majewski, S. R., Guhathakurta, P., et al. 2007, ApJL, 658, 91
\bibitem[Bigiel et al.(2008)]{Bigiel et al. 2008}Bigiel, F., Leroy, A., Walter, F., et al., 2008, AJ, 136, 2846
\bibitem[Bigiel et al.(2011)]{Bigiel et al. 2011}Bigiel, F., Leroy, A. K., Walter, F., et al., 2011, ApJ, 730, L13
\bibitem[Bolatto et al.(2013)]{Bolatto et al. 2013}Bolatto, A. D., Wolfire, M., \& Leroy, A. K. 2013, ARA\&A, 51, 207
\bibitem[Braun et al.(2009)]{Braun et al. 2009} Braun, R., Thilker, D. A., Walterbos, R. A. M., \& Corbelli, E. \ 2009, ApJ, 695, 937
\bibitem[Buckle et al.(2009)]{Buckle 2009}Buckle, J. V., Hills, R. E., Smith, H., et al. \ 2009, MNRAS, 399, 1026
\bibitem[Cald$\rm\acute{u}$-Primo \& Schruba(2016)]{Caldu 2016}Cald$\rm\acute{u}$-Primo, A., \& Schruba, A. 2016, AJ, 151, 34
\bibitem[Ciardullo et al.(1988)]{Ciardullo 1988}Ciardullo, R., Rubin, V. C., Ford, W. K., Jr., Jacoby, G. H., \& Ford, H. C. 1988, AJ, 95, 438
\bibitem[Clark \& Glover(2015)]{Clark 2015}Clark, P. C., \& Glover, S. C. O. 2015, MNRAS, 452, 2057
\bibitem[Currie et al.(2014)]{Currie 2014}Currie M. J., Berry D. S., Jenness T., Gibb A. G., Bell G. S., Draper P. W., 2014, in Manset N., Forshay P., eds, ASP Conf. Ser. Vol. 485, Astronomical Data Analysis Software and Systems XXIII. Astron. Soc. Pac., San Francisco, p. 391
\bibitem[Dalcanton et al.(2012)]{Dalcanton 2012}Dalcanton, J. J., Williams, B. F., Lang, D., et al. 2012, ApJS, 200, 18
\bibitem[de Grijs \& Bono(2014)]{de 2014}de Grijs, R., \& Bono, G. 2014, AJ, 148, 17
\bibitem[Dong et al.(2016)]{Dong et al. 2016} Dong, H., Li, Z., Wang, Q. D., et al. \ 2016, MNRAS, 459, 2262
\bibitem[Draine et al.(2014)]{Draine et al. 2014}Draine B. T., Aniano, G., Krause, O., et al., \ 2014, ApJ, 780, 172
\bibitem[Foreman-Mackey et al.(2013)]{Foreman-Mackey 2013}Foreman-Mackey, D., Hogg, D. W., Lang, D., \& Goodman, J. 2013, PASP, 125, 306
\bibitem[Ford et al.(2013)]{Ford 2013}Ford, G. P., Gear, W. K., Smith, M.W. L. et al. \ 2013, ApJ, 769, 55
\bibitem[Gao \& Solomon(2004)]{Gao 2004}Gao Y., \& Solomon P. M., 2004, ApJ, 606, 271
\bibitem[Goldreich \& Kwan(1974)]{Goldreich 1974}Goldreich, P., \& Kwan, J. \ 1974, ApJ, 189, 441
\bibitem[Gordon et al.(2006)]{Gordon et al. 2006}Gordon, K. D., Bailin, J., Engelbracht, C. W., et al. \ 2006, ApJL, 638, L87
\bibitem[Groves et al.(2012)]{Groves et al. 2012} Groves, B., Krause, O., Sandstrom, K., et al. \ 2012, MNRAS, 426, 892
\bibitem[Jenness \& Economou(2015)]{Jenness 2015}Jenness, T., \& Economou, F. 2015, A\&C, 9, 40
\bibitem[Johnstone et al.(2003)]{Johnstone 2003}Johnstone, D., Boonman, A. M. S., \& van Dishoeck, E. F. 2003, \aap, 412, 157
\bibitem[Kapala et al.(2015)]{Kapala 2015}Kapala, M. J., Sandstrom, K., Groves, B., et al. \ 2015, ApJ, 798, 24
\bibitem[Kennicutt(1989)]{Kennicutt 1989}Kennicutt R. C., \ 1989, ApJ, 344, 685
\bibitem[Kennicutt et al.(2007)]{Kennicutt et al. 2007}Kennicutt, R. C., Calzetti, D., Walter, F., et al., 2007, ApJ, 671, 333
\bibitem[Kennicutt  et al.(2011)]{Kennicutt et al. 2011}Kennicutt, R. C., Calzetti, D., Aniano, G., et al., \ 2011, PASP, 123, 1347
\bibitem[Koda et al.(2012)]{Koda 2012}Koda, J., Scoville, N., Hasegawa, T. et al., \ 2012, ApJ, 761, 41
\bibitem[Komugi et al.(2007)]{Komugi 2007}Komugi, S., Kohno, K., Tosaki, T., et al. 2007, PASJ, 59, 55
\bibitem[Leroy et al.(2008)]{Leroy 2008}Leroy, A. K., Walter, F., Brinks, E., et al. 2008, AJ, 136, 2782
\bibitem[Leroy et al.(2009)]{Leroy et al. 2009}Leroy, A. K., Walter, F., Bigiel, F., et al. \ 2009, AJ, 137, 4670
\bibitem[Lewis et al.(2015)]{Lewis et al. 2015}Lewis, A. R., Dolphin, A. E., Dalcanton, J. J., et al. \ 2015, ApJ, 805, 183
\bibitem[Lewis et al.(2017)]{Lewis 2017}Lewis, A. R., Simones, J. E., Johnson, B. D., et al. 2017, ApJ, 834, 70
\bibitem[Li et al.(2009)]{Li 2009}Li, Z., Wang, Q. D., \& Wakker, B. P., \ 2009, MNRAS, 397, 148
\bibitem[Li et al.(2011)]{Li 2011}Li, Z., Garcia, M. R., Forman, W. R., et al. 2011, ApJ, 728, 10
\bibitem[Li et al.(2019)]{Li et al. 2019} Li, Z., Li, Z., Zhou, P., et al. \ 2019, MNRAS, 484, 964
\bibitem[Loinard \& Allen(1998)]{Loinard 1998}Loinard et al. 1998, ApJ, 499, 227
\bibitem[Loinard et al.(1996)]{Loinard 1996}Loinard et al. 1996, \aap, 310, 93
\bibitem[Mao et al.(2010)]{Mao 2010}Mao, R.-Q., Schulz, A., Henkel, C., et al. 2010, ApJ, 724, 1336
\bibitem[Marsh et al.(2015)]{Marsh 2015}Marsh K. A., Whitworth A. P., Lomax O., 2015, MNRAS, 454, 4282
\bibitem[McConnachie et al.(2005)]{McConnachie 2005}McConnachie, A. W., Irwin, M. J., Ferguson, A. M. N., et al. \ 2005, MNRAS, 356, 979
\bibitem[McMullin et al.(2007)]{McMullin 2007}McMullin, J. P., Waters, B., Schiebel, D., Young, W., \& Golap, K. 2007, in ASP Conf. Ser. 376, Astronomical Data Analysis Software and Systems XVI, ed. R. A. Shaw, F. Hill, \& D. J. Bell (San Francisco, CA: ASP), 127
\bibitem[Melchior \& Combes(2011)]{Melchior 2011} Melchior, A.~L. \& Combes, F. \ 2011, \aap, 536, A52
\bibitem[Montalto et al.(2009)]{Montalto et al. 2009}Montalto, M., Seitz, S., Riffeser, A., et al. 2009, \aap, 507, 283
\bibitem[Muraoka et al.(2007)]{Muraoka et al. 2007}Muraoka, K., Kohno, K., Tosaki, T., et al., 2007, PASJ, 59, 43
\bibitem[Muraoka et al.(2016)]{Muraoka 2016}Muraoka, K., Takeda, M., Yanagitani, K., et al. \ 2016, PASJ, 68, 18
\bibitem[Nieten et al.(2006)]{Nieten}Nieten Ch., Neininger N., Gu$\rm{\acute{e}}$lin M., Ungerechts H., et al. \ 2006, A\&A, 453, 459
\bibitem[Oka et al.(2007)]{Oka et al. 2007}Oka, T., Nagai, M., Kamegai, K., Tanaka, K., \& Kuboi, N. \ 2007, PASJ, 59, 15
\bibitem[Oka et al.(2012)]{Oka et al. 2012} Oka, T., Onodera, Y., Nagai, M., et al. \ 2012, ApJS, 201, 14
\bibitem[Querejeta et al.(2019)]{Querejeta 2019}Querejeta M., et al., 2019, A\&A, 625, A19
\bibitem[Rahmani et al.(2016)]{Rahmani 2016}Rahmani, S., Lianou, S., \& Barmby, P. 2016, MNRAS, 456, 4128
\bibitem[Saglia et al.(2018)]{Saglia 2018}Saglia, R. P., Opitsch, M., Fabricius, M. H., et al. 2018, A\&A, 618, A156
\bibitem[Saintonge et al.(2017)]{Saintonge 2017}Saintonge, A., Catinella, B., Tacconi, L. J., et al. 2017, ApJS, 233, 22
\bibitem[Sanders et al.(1993)]{Sanders et al. 1993}Sanders, D. B., Scoville, N. Z., Tilanus, R. P. J., Wang, Z., \& Zhou, S. \ 1993, Back to the Galaxy, 278, 311
\bibitem[Schinnerer et al.(2010)]{Schinnerer 2010}Schinnerer, E., Wei$\beta$, A., Aalto, S., \& Scoville, N. Z. \ 2010, ApJ, 719, 1588
\bibitem[Schmidt(1959)]{Schmidt 1959}Schmidt, M. 1959, ApJ, 129, 243
\bibitem[Scoville \& Sanders(1987)]{Scoville 1987}Scoville, N. Z., \& Sanders, D. B. 1987, in Astrophysics and Space Science Library, Vol. 134, Interstellar Processes, ed. D. J. Hollenbach \& H. A. Thronson, Jr. (Dordrecht: Reidel), 21
\bibitem[Scoville \& Solomon(1974)]{Scoville 1974} Scoville, N. Z., \& Solomon, P. M. \ 1974, ApJ, 187, L67
\bibitem[Shetty et al.(2011)]{Shetty 2011}Shetty, R., Glover, S. C., Dullemond, C. P., Ostriker, E. C., Harris, A. I., \& Klessen, R. S. 2011, MNRAS, 415, 3253
\bibitem[Shetty et al.(2013)]{Shetty 2013}Shetty, R., Kelly, B. C., \& Bigiel, F. 2013, MNRAS, 430, 288
\bibitem[Shi et al.(2018)]{Shi 2018}Shi, Y., Yan, L., Armus, L., et al. 2018, ApJ, 853, 149
\bibitem[Smith et al.(2012)]{Smith 2012} Smith, M. W. L., Eales, S. A., Gomez,  H. L., et al. \ 2012, ApJ, 756, 40
\bibitem[Solomon et al.(1987)]{Solomon 1987}Solomon, P. M., Rivolo, A. R., Barrett, J., \& Yahil, A. 1987, ApJ, 319, 730
\bibitem[Sun et al.(2018)]{Sun 2018}Sun, J., Leroy, A. K., Schruba, A. et al., \ 2018, ApJ, 860, 172
\bibitem[Tabatabaei \& Berkhuijsen(2010)]{Tabatabaei 2010}Tabatabaei, F. S., \& Berkhuijsen, E. M. 2010, \aap, 517, A77
\bibitem[Thilker et al.(2005)]{Thilker et al. 2005}Thilker, D. A., Hoopes, C. G., Bianchi, L., et al. \ 2005, ApJL, 619, L67
\bibitem[Tomi$\rm \check{c}i\acute{c}$ et al.(2019)]{Tomicic 2019}Tomi$\rm \check{c}i\acute{c}$, N., Ho, I.-T., Kreckel, K., et al. 2019, ApJ, 873, 3
\bibitem[van der Tak et al.(2007)]{van 2007} van der Tak, F. F. S., Black, J. H., Sch$\rm \ddot{o}$ier, F. L., Jansen, D. J., \& van Dishoeck, E. F. \ 2007, \aap, 468, 627
\bibitem[Viaene et al.(2017)]{Viaene et al. 2017}Viaene, S., Baes, M., Tamm, A., et al., 2017, \aap, 599, A64
\bibitem[Viaene et al.(2018)]{Viaene 2018}Viaene, S., Forbrich, J., \& Fritz, J. 2018, MNRAS, 475, 5550
\bibitem[Vlahakis et al.(2013)]{Vlahakis et al. 2013}Vlahakis, C., van der Werf, P., Israel, F. P., \& Tilanus, R. P. J. \ 2013, MNRAS, 433, 1837
\bibitem[Walter et al.(2008)]{Walter et al. 2008}Walter F., Brinks E., de Blok W. J. G., Bigiel F., Kennicutt R. C., Thornley M. D., Leroy A. K., 2008, AJ, 136, 2563
\bibitem[Watson(2010)]{Watson 2010}Watson M. E., 2010, Master's thesis, University of Hertfordshire
\bibitem[Whitworth et al.(2019)]{Whitworth 2019}Whitworth A. P., Marsh K. A., Cigan P. J., et al. 2019, ArXiv: 1908.03458
\bibitem[Williams et al.(1994)]{Williams 1994}Williams J. P., de Geus E. J., Blitz L., 1994, ApJ, 428, 693
\bibitem[Williams et al.(2018)]{Williams 2018}Williams T. G., Gear W. K., \& Smith M. W. L., 2018, MNRAS, 479, 297
\bibitem[Wilson \& Scoville(1989)]{Wilson 1989}Wilson, C. D., \& Scoville, N., 1989, ApJ, 347, 743
\bibitem[Wilson et al.(1997)]{Wilson et al. 1997}Wilson C. D., Walker C. E., Thornley M. D., 1997, ApJ, 483, 210
\bibitem[Wilson et al.(2009)]{Wilson 2009}Wilson, C. D., Warren, B. E., Israel, F. P., et al. \ 2009, ApJ, 693, 1736
\bibitem[Wilson et al.(2012)]{Wilson 2012}Wilson, C. D., Warren, B. E., Israel, F. P., et al. \ 2012, MNRAS, 424, 3050

\end{thebibliography}
\end{document}